\documentclass[aps,preprint]{revtex4}%
\usepackage{amsfonts}
\usepackage{amsmath}
\usepackage{amssymb}
\usepackage[colorlinks,linkcolor=blue,citecolor=blue,urlcolor=blue]{hyperref}
\usepackage{subfigure}
\usepackage{graphicx}
\usepackage{cleveref}
\usepackage{listings}
\usepackage{xcolor}
\usepackage{array}
\usepackage{url}%
\begin{document}
\preprint{CTP-SCU/2020021}
\title{Extended Phase Space Thermodynamics for Black Holes in a Cavity}
\author{Peng Wang}
\email{pengw@scu.edu.cn}
\author{Houwen Wu}
\email{iverwu@scu.edu.cn}
\author{Haitang Yang}
\email{hyanga@scu.edu.cn}
\author{Feiyu Yao}
\email{yaofeiyu@stu.scu.edu.cn}
\affiliation{Center for Theoretical Physics, College of Physics, Sichuan University,
Chengdu, 610064, China}

\begin{abstract}
In this paper, we extend the phase space of black holes enclosed by a
spherical cavity of radius $r_{B}$ to include $V\equiv4\pi r_{B}^{3}/3$ as a
thermodynamic volume. The thermodynamic behavior of Schwarzschild and
Reissner-Nordstrom (RN) black holes is then investigated in the extended phase
space. In a canonical ensemble at constant pressure, we find that the
aforementioned thermodynamic behavior is remarkably similar to that of the
anti-de Sitter (AdS) counterparts with the cosmological constant being
interpreted as a pressure. Specifically, a first-order Hawking-Page-like phase
transition occurs for a Schwarzschild black hole in a cavity. The phase
structure of a RN black hole in a cavity shows a strong resemblance to that of
the van der Waals fluid. Our results may provide a new perspective for the
extended thermodynamics of AdS black holes by analogy with black holes in a cavity.

\end{abstract}
\keywords{}\maketitle
\tableofcontents

\section{Introduction}

The area theorem of black holes \cite{Hawking:1971tu} asserts that the total
horizon area of black holes is a non-decreasing function of time in reasonable
physical processes, which hints that black holes are endowed with
thermodynamic properties. Since the area law bears a close resemblance to the
second law of thermodynamics, Bekenstein postulated that black hole entropy is
proportional to the horizon area \cite{Bekenstein:1972tm,Bekenstein:1973ur}.
The analogy between usual thermodynamics and black hole thermodynamics was
further confirmed by the discovery of Hawking radiation, assigning black holes
a temperature \cite{Hawking:1974rv,Hawking:1975iha}. Analogous to the laws of
thermodynamics, the four laws of black hole mechanics were established in
\cite{Bardeen:1973gs}.

Partly due to AdS/CFT correspondence \cite{Maldacena:1997re}, nearly fifty
years after the discovery of black hole thermodynamics, understanding
thermodynamic properties of various black holes, especially phase transitions
of AdS black holes, is still a rather hot topic in the literature. Since AdS
boundary plays a role of a reflecting wall, AdS black holes can be thermally
stable in some cases, which makes it possible to study their phase behavior.
In the seminal work \cite{Hawking:1982dh}, the Hawking-Page phase transition
(i.e., a phase transition between the thermal AdS space and a black hole) was
revealed in Schwarzschild-AdS black holes. Subsequently, there has been much
interest in studying thermodynamics and phase structure of AdS black holes
\cite{Witten:1998zw,Cvetic:1999ne,Chamblin:1999tk,Chamblin:1999hg,Caldarelli:1999xj,Cai:2001dz,Cvetic:2001bk,Nojiri:2001aj}%
. Interestingly, RN-AdS black holes were found to possess a van der Waals-like
phase transition (i.e., a phase transition consisting of a first-order phase
transition terminating at a second-order critical point) in a canonical
ensemble \cite{Chamblin:1999tk,Chamblin:1999hg} and a Hawking-Page-like phase
transition in a grand canonical ensemble \cite{Peca:1998cs}.

Later, the cosmological constant in AdS black holes is identified as a
thermodynamic pressure \cite{Dolan:2011xt,Kubiznak:2012wp}. In this framework,
the first law becomes consistent with the corresponding Smarr relation, and
black hole mass is interpreted as a chemical enthalpy \cite{Kastor:2009wy}.
The phase behavior and $P$-$V$ criticality have been explored for various AdS
black holes in the extended phase space
\cite{Wei:2012ui,Gunasekaran:2012dq,Cai:2013qga,Xu:2014kwa,Frassino:2014pha,Dehghani:2014caa,Hennigar:2015esa,Caceres:2015vsa,Hendi:2016yof,Hendi:2017fxp,Lemos:2018cfd,Pedraza:2018eey,Wang:2018xdz,Wei:2020poh}%
, which discovered a broad range of new phenomena. For a recent review, see
\cite{Kubiznak:2016qmn}. Specifically for a Schwarzschild-AdS black hole, the
coexistence line of the Hawking-Page phase transition in the $P$-$T$ diagram
is semi-infinite and reminiscent of the solid/liquid phase transition
\cite{Kubiznak:2014zwa}. In the extended phase space, the analogy between a
RN-AdS black hole and the van der Waals fluid becomes more complete, in that
the coexistence lines in the $P$-$T$ diagram are both finite and terminate at
critical points, and the $P$-$V$ criticality matches with one another
\cite{Kubiznak:2012wp}.

In parallel with research on AdS black holes, studies of thermodynamics of
black holes in a cavity have also attracted a lot attentions since York
realized that Schwarzschild black holes can be thermally stable by placing
them inside a spherical cavity, on the wall of which the metric is fixed
\cite{York:1986it}. Phase structure and transitions of Schwarzschild black
holes in a cavity and Schwarzschild-AdS black holes were shown to be
strikingly similar \cite{York:1986it}. Afterwards, it was found that the phase
behavior of RN black holes in a cavity and RN-AdS black holes also has
extensive similarities in a grand canonical ensemble \cite{Braden:1990hw} and
a canonical ensemble \cite{Carlip:2003ne,Lundgren:2006kt}. Similar analysis
has been extended to a broad class of brane backgrounds, including $Dp$-brane
and NS$5$-brane configurations, and most of the brane systems were observed to
undergo Hawking-Page-like or van der Waals-like phase transitions
\cite{Lu:2010xt,Wu:2011yu,Lu:2012rm,Lu:2013nt,Zhou:2015yxa,Xiao:2015bha}. In
addition, properties of boson stars and hairy black holes in a cavity were
investigated
\cite{Sanchis-Gual:2015lje,Sanchis-Gual:2016tcm,Basu:2016srp,Peng:2017gss,Peng:2017squ,Peng:2018abh,Kiczek:2019qbk,Kiczek:2020gyd}%
, which were shown to closely resemble those of holographic superconductors in
the AdS gravity. Lately, it was discovered that Gauss-Bonnet black holes in a
cavity have quite similar phase structure and transitions to the AdS
counterparts \cite{Wang:2019urm}. These observations lend support to the
analogy between black holes in a cavity and AdS black holes.

However, we recently studied Born-Infeld black holes enclosed in a cavity in a
canonical ensemble \cite{Wang:2019kxp} and a grand canonical ensemble
\cite{Liang:2019dni} and revealed that their phase structure has
dissimilarities from that of Born-Infeld-AdS black holes. Moreover, contrary
to the phase behavior, we found that there exist significant differences
between the thermodynamic geometry of RN black holes in a cavity and that of
RN-AdS black holes \cite{Wang:2019cax}, and some dissimilarities between the
two cases also occur for validities of the second thermodynamic law and the
weak cosmic censorship \cite{Wang:2020osg}. These findings motivate us to
further explore connections between thermodynamic properties of black holes
and their boundary conditions. Additionally, it has been proposed
\cite{McGough:2016lol} that the holographic dual of $T\bar{T}$ deformed
CFT$_{\text{2}}$ is a finite region of AdS$_{\text{3}}$ with the wall at
finite radial distance, which makes studying properties of black holes in a
cavity more attractive. Note that thermodynamics and critical behavior of de
Sitter black holes in a cavity were recently investigated in
\cite{Simovic:2018tdy,Simovic:2019zgb,Haroon:2020vpr}.

In the existing research on thermodynamic properties of asymptotically flat
black holes surrounded by a cavity, thermodynamic quantities that can be
interpreted as a pressure or volume are absent. To make analogies with the AdS
counterparts and corresponding real-world systems more complete, it is highly
desirable to introduce a thermodynamic pressure or volume for black holes in a
cavity. To this end, we extend the phase space to include the cavity radius as
a new thermodynamic variable, from which the thermodynamic pressure and volume
of black holes in a cavity can be established. Particularly in this paper, we
confine Schwarzschild and RN black holes in a cavity and study their
thermodynamic behavior in the aforementioned extended phase space. In
hindsight, the extended phase behavior of Schwarzschild and RN black holes in
a cavity is discovered to bear a striking resemblance to that of the AdS counterparts.

The rest of this paper is organized as follows. Section \ref{Sec:SBHC}
discusses phase structure of Schwarzschild black holes in a cavity in the
extended phase space. In section \ref{Sec:RNBHC}, we investigate the extended
phase properties of RN black holes in a cavity in a canonical ensemble that
maintains constant temperature, pressure and charge. We summarize our results
with a brief discussion in section \ref{Sec:DC}. For simplicity, we set
$G=\hbar=c=k_{B}=1$ in this paper.

\section{Schwarzschild Black Hole in a Cavity}

\label{Sec:SBHC}

In this section, we consider a thermodynamic system with a Schwarzschild black
hole enclosed in a cavity. The $4$-dimensional Schwarzschild black hole
solution is described by%
\begin{equation}
ds^{2}=-f\left(  r\right)  dt^{2}+\frac{dr^{2}}{f\left(  r\right)  }%
+r^{2}\left(  d\theta^{2}+\sin^{2}\theta d\phi^{2}\right)  ,\text{ }f\left(
r\right)  =1-\frac{r_{+}}{r},
\end{equation}
where $r_{+}$ is the radius of event horizon. The Hawking temperature $T_{b}$
of the Schwarzschild black hole is%
\begin{equation}
T_{b}=\frac{1}{4\pi r_{+}}.
\end{equation}
Suppose the wall of the cavity enclosing the black hole is at $r=r_{B}\geq
r_{+}$ and maintained at a temperature of $T.$ In \cite{Braden:1990hw}, it
showed that $T$ can be related to $T_{b}$ as%
\begin{equation}
T=\frac{T_{b}}{\sqrt{f\left(  r_{B}\right)  }}. \label{eq:hwT}%
\end{equation}
The thermal energy $E$ and the Helmholtz free energy $F$ were derived in
\cite{Carlip:2003ne}:%
\begin{align}
E  &  =r_{B}\left[  1-\sqrt{f\left(  r_{B}\right)  }\right]  ,\nonumber\\
F  &  =r_{B}\left[  1-\sqrt{f\left(  r_{B}\right)  }\right]  -T\pi r_{+}.
\label{eq:F}%
\end{align}

We now introduce a new thermodynamic variable, the thermodynamic volume of the
system $V$,%
\begin{equation}
V\equiv\frac{4}{3}\pi r_{B}^{3}, \label{eq:V}%
\end{equation}
which naturally gives the conjugate thermodynamic pressure,%
\begin{equation}
P=-\frac{\partial E}{\partial V}=-\frac{1}{4\pi r_{B}^{2}}\left(  1-\sqrt
{1-x}-\frac{x}{2\sqrt{1-x}}\right)  ,\text{ }0\leq x\equiv\frac{r_{+}}{r_{B}%
}\leq1. \label{eq:SchP}%
\end{equation}
Extending the phase space of black hole thermodynamics to include the
pressure/volume conjugate pair leads to the first law of thermodynamics,%
\begin{equation}
dE=TdS-PdV,
\end{equation}
where $S=\pi r_{+}^{2}$ is the entropy of black hole, and we use $T=\partial
E/\partial S$. If the system undergoes an isobaric process, the Gibbs free
energy $G\equiv F+PV$ is employed in the study of phases and phase transitions.

\begin{figure}[tb]
\begin{center}
\includegraphics[width=0.48\textwidth]{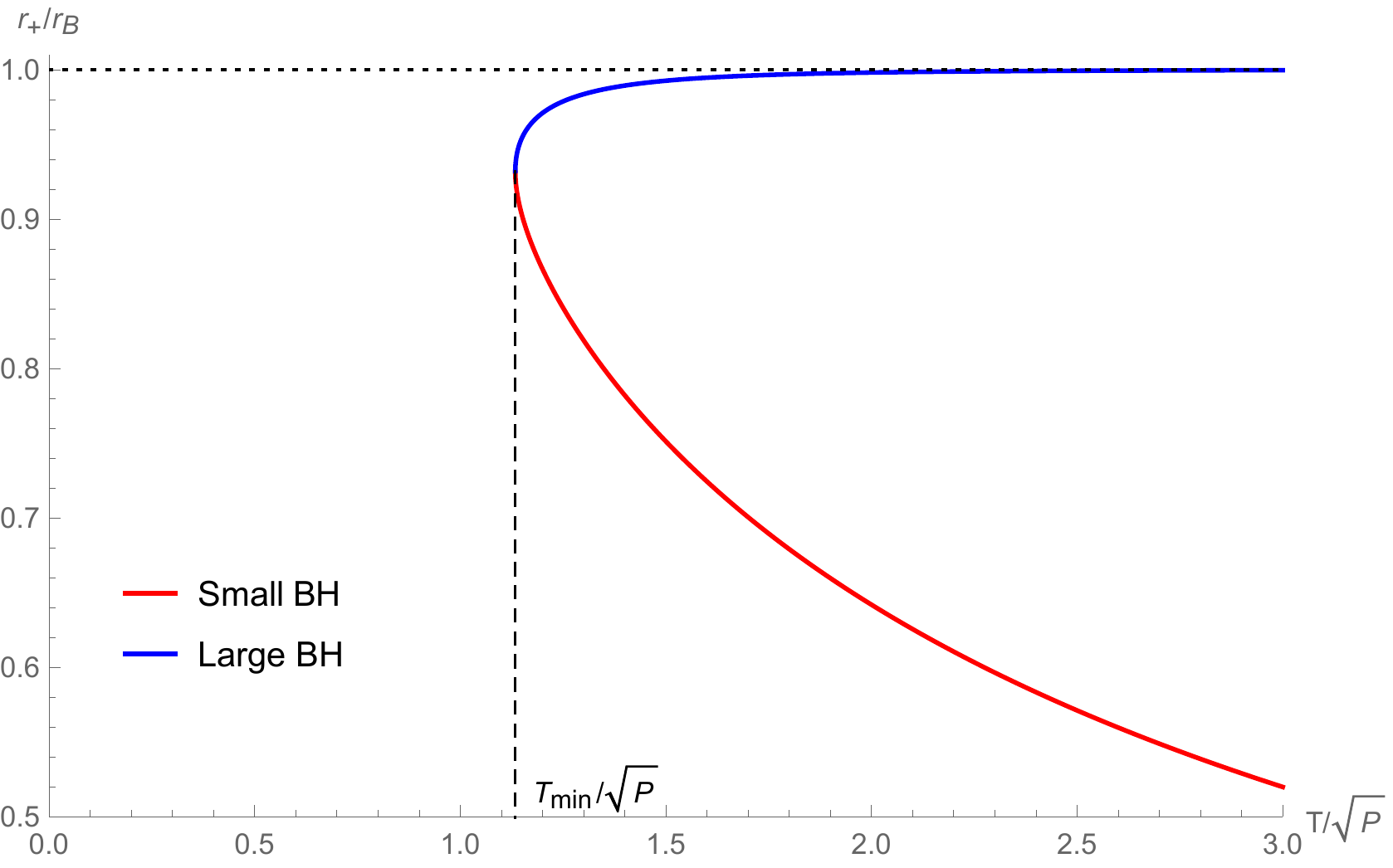}\label{fig:Schr:a}
\includegraphics[width=0.48\textwidth]{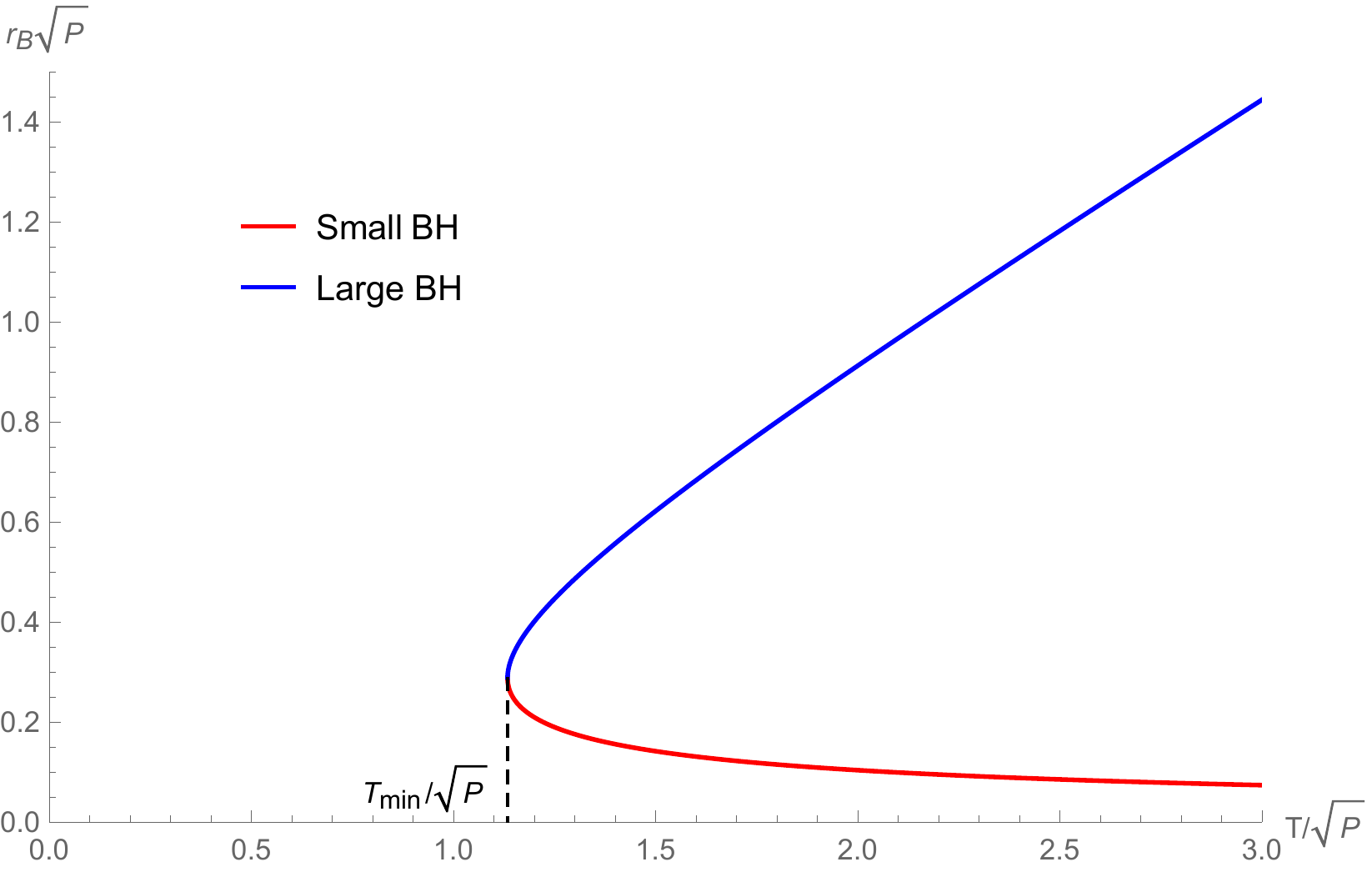}\label{fig:Schr:b}
\end{center}
\caption{{\footnotesize Plots of $r_{+}/r_{B}\ $and $r_{B}\sqrt{P}$\ against
$T/\sqrt{P}$\ for a Schwarzschild black hole in a cavity. The red and blue
lines represent Small BH and Large BH, respectively. Black hole solutions do
not exist when $T<T_{\min}$.}}%
\label{fig:Schr}%
\end{figure}

By dimensional analysis, we find that the thermodynamic quantities scale as
powers of the pressure $P$,%
\begin{equation}
T=\tilde{T}\sqrt{P}\text{, }G=\tilde{G}/\sqrt{P}\text{, }r_{B}=\tilde{r}%
_{B}/\sqrt{P}\text{,}%
\end{equation}
where the tildes denote dimensionless quantities. Solving eqn. $\left(
\ref{eq:hwT}\right)  $ for $x$ in terms of $\tilde{T}$ gives $x=x(\tilde{T})$,
which is plotted in the left panel of FIG. \ref{fig:Schr}. It shows that for
$T\geq T_{\min}$, $x=x(\tilde{T})$ is multivalued and consists of two
branches, namely Small BH and Large BH. When $T<T_{\min}$, no black hole
exists. Plugging $x(\tilde{T})$ into eqn. $\left(  \ref{eq:SchP}\right)  $,
one can express $\tilde{r}_{B}$ in terms of $\tilde{T}$: $\tilde{r}_{B}%
=\tilde{r}_{B}(\tilde{T})$. In the right panel of FIG. \ref{fig:Schr}, we show
that $\tilde{r}_{B}(\tilde{T})$ is also composed of the Small BH and Large BH
branches. Note that the Small (Large) BH branch possesses smaller (larger)
$r_{+}$ and $r_{B}$.

\begin{figure}[tb]
\begin{center}
\includegraphics[width=0.48\textwidth]{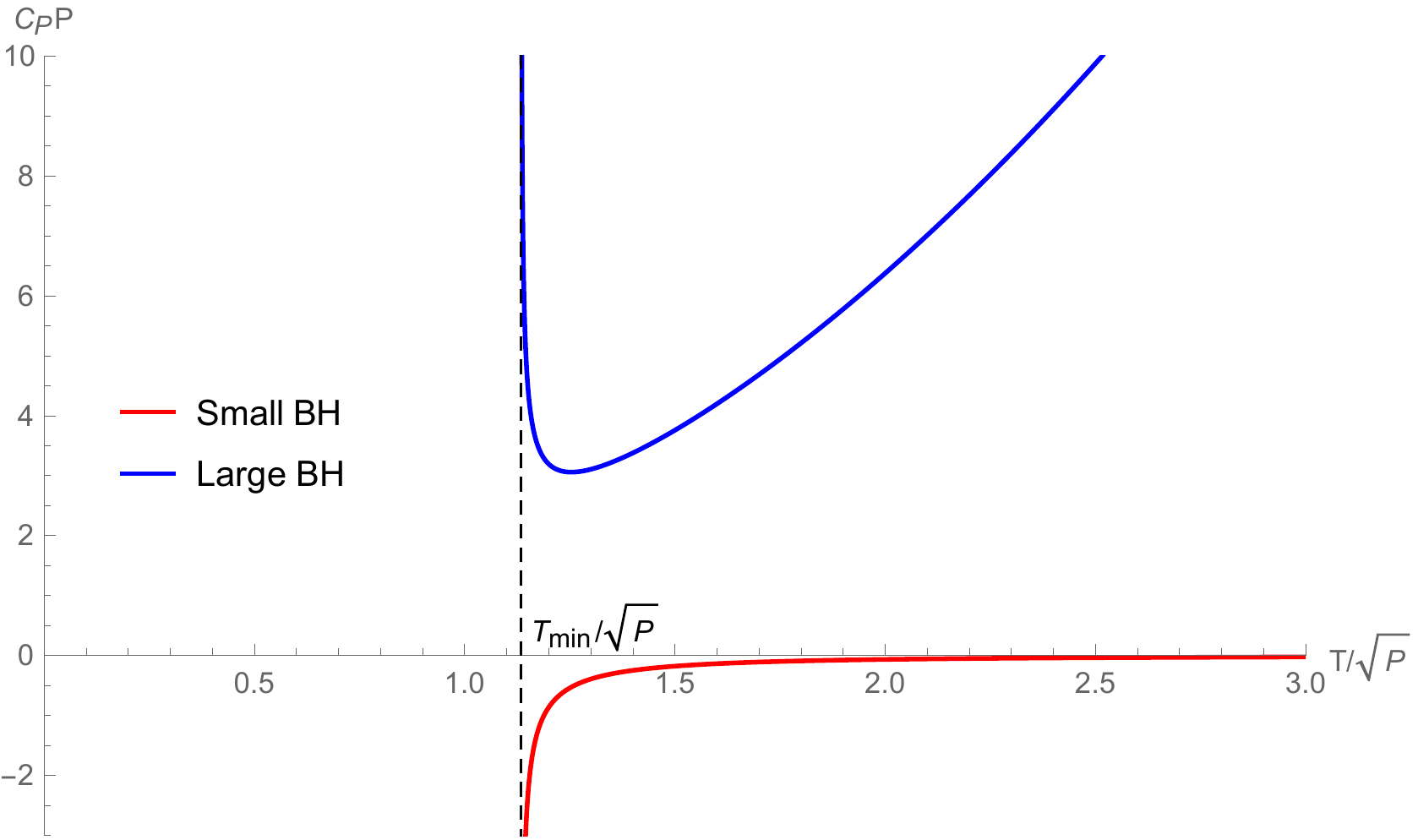}\label{fig:Schp:a}
\includegraphics[width=0.48\textwidth]{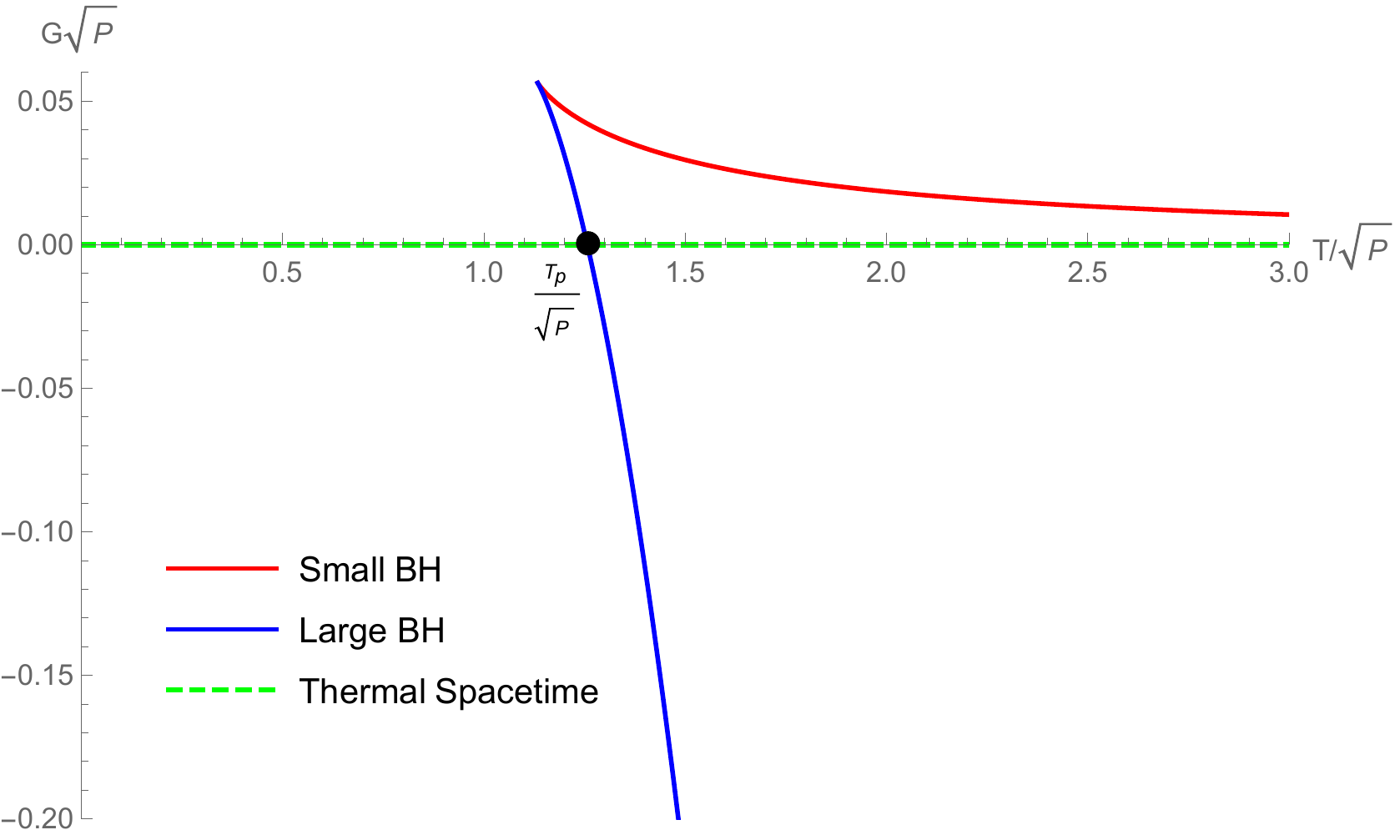}\label{fig:Schp:b}
\end{center}
\caption{{\footnotesize Plots of $C_{p}P$ and $G\sqrt{P}$\ against $T/\sqrt
{P}$\ for a Schwarzschild black hole in a cavity. \textbf{Left Panel: }The
heat capacity at constant pressure $C_{p}$ of Small/Large BH is
negative/positive, which means that Small/Large BH is thermally
unstable/stable in an isobaric process. \textbf{Right Panel: }As $T/\sqrt{P}$
increases from zero, a first-order phase transition from the thermal spacetime
to Large BH occurs at the black dot. Above $T_{p}/\sqrt{P}$, Large BH is the
thermodynamically preferred state.}}%
\label{fig:Schp}%
\end{figure}

To study the thermodynamic stability of the two branches against thermal
fluctuations in an isobaric process, we consider the heat capacity at constant
pressure,%
\begin{equation}
C_{P}=\tilde{C}_{P}/P=T\left(  \frac{\partial S}{\partial T}\right)  _{P}.
\end{equation}
Using $x(\tilde{T})$, we can rewrite $\tilde{C}_{P}$ as%
\begin{equation}
\tilde{C}_{P}=\tilde{C}_{P}(\tilde{T}),
\end{equation}
which is presented in the left panel of FIG. \ref{fig:Schp}. It shows that
Small (Large) BH is thermally unstable (stable). To study phase transitions,
we also need to consider the thermal flat spacetime as a phase of the system
since it is a classical solution in the canonical ensemble. The Gibbs free
energy $\tilde{G}(\tilde{T})$ of the two branches and the thermal flat
spacetime is displayed in the right panel of FIG. \ref{fig:Schp}. One finds
that the thermal spacetime is the only phase when $\tilde{T}<\tilde{T}_{\min}%
$, and the black hole appears when $\tilde{T}\geq\tilde{T}_{\min}$. Similar to
a Schwarzschild-AdS black hole, a first-order Hawking-Page-like phase
transition between the thermal spacetime and Large BH occurs at $\tilde
{T}=\tilde{T}_{p}$, where these two phases are of equal Gibbs free energy. The
thermal spacetime is globally stable for $\tilde{T}<\tilde{T}_{p}$ while Large
BH is globally preferred for $\tilde{T}\geq\tilde{T}_{p}$. The coexistence
line of thermal spacetime/Large BH phases is determined by $\tilde{G}=0$,
which reads%
\begin{equation}
T|_{\text{coexistence}}\simeq1.257\sqrt{P}.
\end{equation}
The coexistence line is of infinite length and hence reminiscent of the
solid/liquid phase transition. It is noteworthy that the coexistence line of
thermal AdS spacetime/Large Schwarzschild-AdS BH was given by
\cite{Kubiznak:2016qmn}%
\begin{equation}
T|_{\text{coexistence}}\simeq0.921\sqrt{P}.
\end{equation}

\begin{figure}[tb]
\begin{center}
\includegraphics[width=0.45\textwidth]{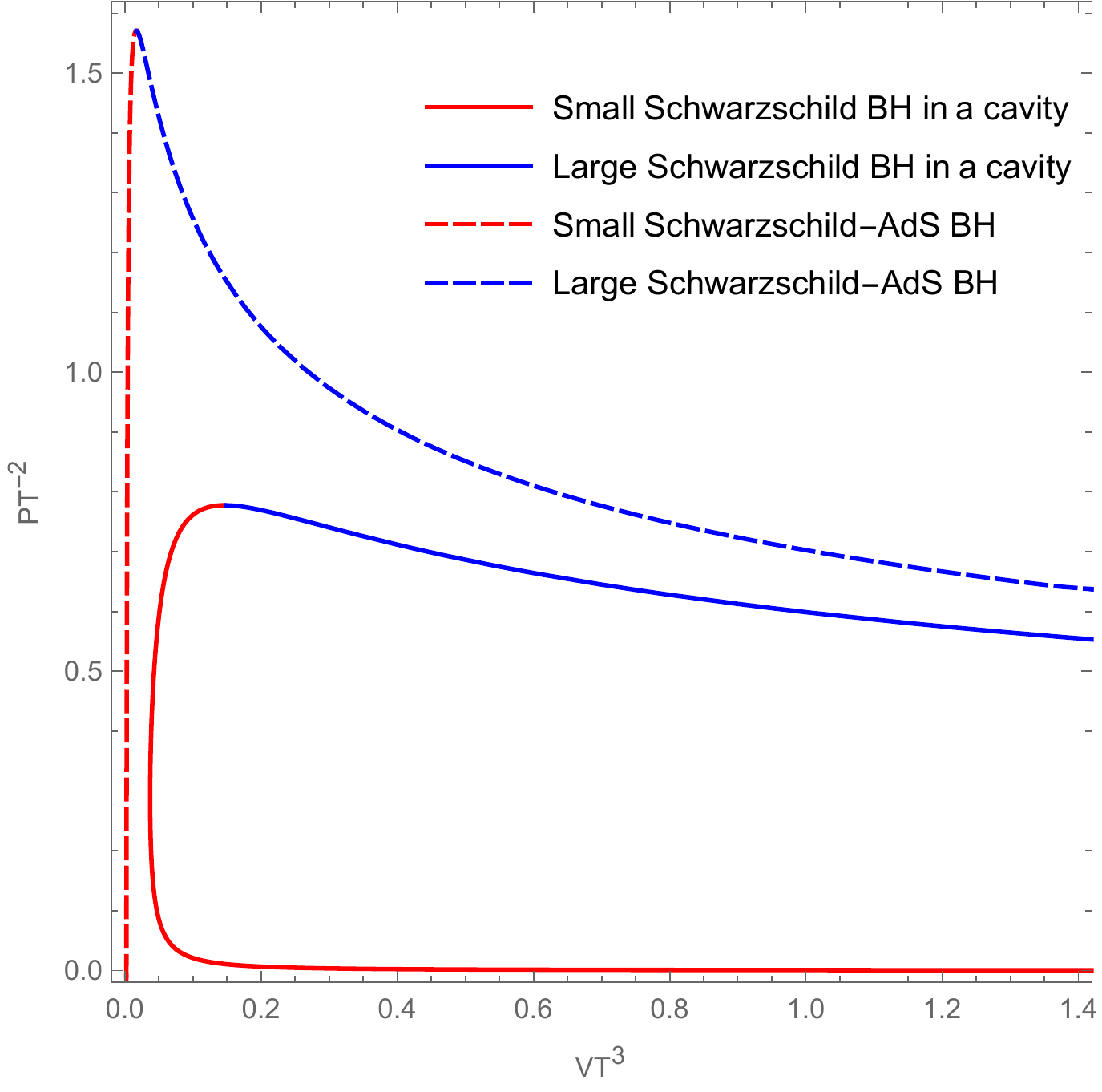}
\end{center}
\caption{{\footnotesize Plot of the equations of state for a Schwarzschild
black hole in a cavity and a Schwarzschild-AdS black hole.}}%
\label{fig:SchEoS}%
\end{figure}

Finally, we discuss the equation of state for the Schwarzschild black hole in
a cavity. By rewriting the pressure equation $\left(  \ref{eq:SchP}\right)  $
whilst using the temperature equation $\left(  \ref{eq:hwT}\right)  $, we can
obtain the equation of state in terms of $VT^{3}$ and $PT^{-2}$, which is
depicted as solid lines in FIG. \ref{fig:SchEoS}. With fixed $PT^{-2}$, the
equation of state has two branches, corresponding to Small BH and Large BH. In
the limit of large $VT^{3}$, the equation of state becomes
\begin{align}
VT^{3}  &  \simeq\frac{1}{48\pi^{2}}\left(  \frac{\pi}{2PT^{-2}}\right)
^{3/4}\text{ for Small BH,}\nonumber\\
VT^{3}  &  \simeq\frac{1}{48\pi^{2}}\left(  \frac{2\pi}{PT^{-2}}\right)
^{3}\text{ \ \ for Large BH.} \label{eq:SchEoSL}%
\end{align}
Defining a specific volume $v\equiv\left(  6V/\pi\right)  ^{1/3}$, the Large
BH equation of state in eqn. $\left(  \ref{eq:SchEoSL}\right)  $ gives the
ideal gas law, $Pv\simeq T$. We also plot the equation of state for a
Schwarzschild-AdS black hole, which is represented by dashed lines in FIG.
\ref{fig:SchEoS}. Similarly, there exist two branches (i.e., Small BH and
Large BH) for a fixed value of $PT^{-2}$. However, only Large BH exists when
$VT^{3}$ is large enough. It is worth noting that, in the limit of large
$VT^{3}$, the equation of state for the Schwarzschild-AdS black hole reduces
to \cite{Kubiznak:2016qmn}%
\begin{equation}
VT^{3}\simeq\frac{1}{48\pi^{2}}\left(  \frac{2\pi}{PT^{-2}}\right)  ^{3}\text{
,}%
\end{equation}
which is the same as that for the Large BH branch of the Schwarzschild black
hole in a cavity.

\section{RN Black Hole in a Cavity}

\label{Sec:RNBHC}

In this section, we discuss phase structure and transitions of a RN black hole
in a cavity in the extended phase space. The $4$-dimensional RN black hole
solution is%
\begin{align}
ds^{2}  &  =-f\left(  r\right)  dt^{2}+\frac{dr^{2}}{f\left(  r\right)
}+r^{2}\left(  d\theta^{2}+\sin^{2}\theta d\phi^{2}\right)  \text{,
}\nonumber\\
f\left(  r\right)   &  =\left(  1-\frac{r_{+}}{r}\right)  \left(
1-\frac{Q_{b}^{2}}{r_{+}r}\right)  \text{, }A=A_{t}\left(  r\right)
dt=-\frac{Q_{b}}{r}dt,
\end{align}
where $Q_{b}$ is the black hole charge, and $r_{+}$ is the radius of the outer
event horizon. The Hawking temperature $T_{b}$ of the RN black hole is given
by%
\begin{equation}
T_{b}=\frac{1}{4\pi r_{+}}\left(  1-\frac{Q_{b}^{2}}{r_{+}^{2}}\right)  .
\end{equation}
In a canonical ensemble, the wall of the cavity, which is located at $r=r_{B}%
$, is maintained at a temperature of $T$ and a charge of $Q$. It was showed in
\cite{Braden:1990hw} that the system temperature $T$ and charge $Q$ can be
related to the black hole temperature $T_{b}$ and charge $Q_{b}$ as%
\begin{equation}
Q=Q_{b}\text{ and }T=\frac{T_{b}}{\sqrt{f\left(  r_{B}\right)  }},
\label{eq:RNTQ}%
\end{equation}
respectively. For this system, the Helmholtz free energy $F$ and the thermal
energy $E$ were also given in \cite{Carlip:2003ne}
\begin{align}
F  &  =r_{B}\left[  1-\sqrt{f\left(  r_{B}\right)  }\right]  -\pi Tr_{+}%
^{2},\nonumber\\
E  &  =r_{B}\left[  1-\sqrt{f\left(  r_{B}\right)  }\right]  .
\end{align}
The physical range of $r_{+}$ is constrained by%
\begin{equation}
\frac{r_{e}}{r_{B}}\leq x\equiv\frac{r_{+}}{r_{B}}\leq1\text{,}
\label{eq:rBound}%
\end{equation}
where $r_{e}=Q$ is the horizon radius of the extremal black hole.

By introducing $V\equiv4\pi r_{B}^{3}/3$ as a thermodynamic variable, the
pressure of the system is
\begin{equation}
P=-\frac{\partial E}{\partial V}=\frac{2r_{+}r_{B}-Q^{2}-r_{+}^{2}}{8\pi
r_{B}^{3}r_{+}\sqrt{\left(  1-\frac{Q^{2}}{r_{+}r_{B}}\right)  \left(
1-\frac{r_{+}}{r_{B}}\right)  }}-\frac{1}{4\pi r_{B}^{2}}. \label{eq:RNP}%
\end{equation}
In this extend phase space, the first law of thermodynamics becomes%
\begin{equation}
dE=TdS-PdV+\Phi dQ,
\end{equation}
where the system's potential $\Phi$ is defined as%
\begin{equation}
\Phi\equiv\frac{A_{t}\left(  r_{B}\right)  -A_{t}\left(  r_{+}\right)  }%
{\sqrt{f\left(  r_{B}\right)  }}.
\end{equation}
A system under constant pressure in the canonical ensemble is best described
by the Gibbs free energy, $G=F+PV$. Two phases in equilibrium have equal Gibbs
free energy. Moreover, the heat capacity at constant pressure $C_{P}=T\left(
\partial S/\partial T\right)  _{P}$ is a thermodynamic quantity measuring the
stability of a phase in an isobaric process. For later convenience, we can
express the thermodynamic quantities in units of $\sqrt{P}$%
\begin{equation}
\tilde{Q}\equiv Q\sqrt{P},\text{ }\tilde{r}_{B}\equiv r_{B}\sqrt{P},\tilde
{T}\equiv T/\sqrt{P},\tilde{G}\equiv G\sqrt{P},\tilde{C}_{P}\equiv C_{P}P.
\end{equation}

\begin{figure}[tb]
\begin{center}
\includegraphics[width=0.48\textwidth]{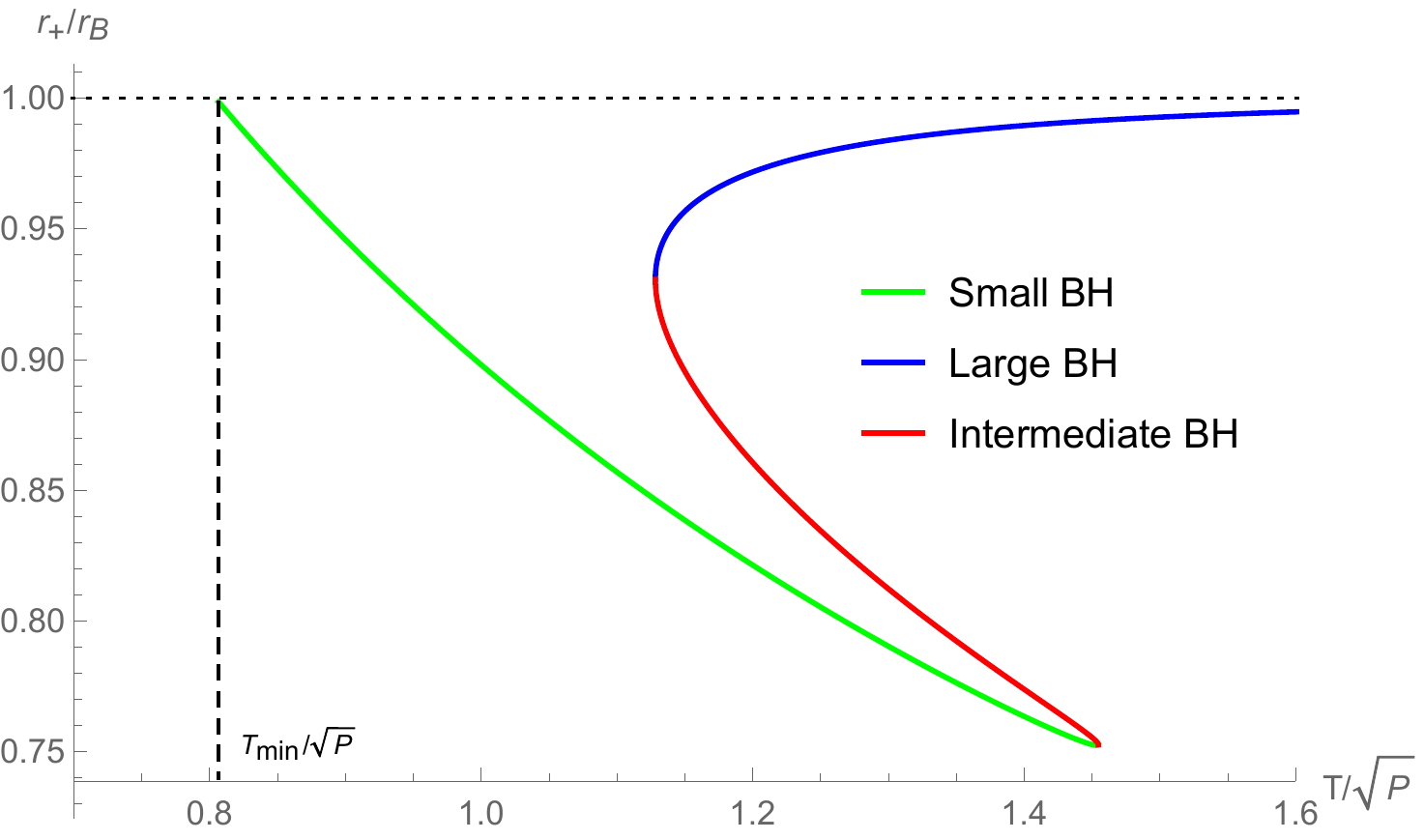}\label{fig:RNrS:a}
\includegraphics[width=0.48\textwidth]{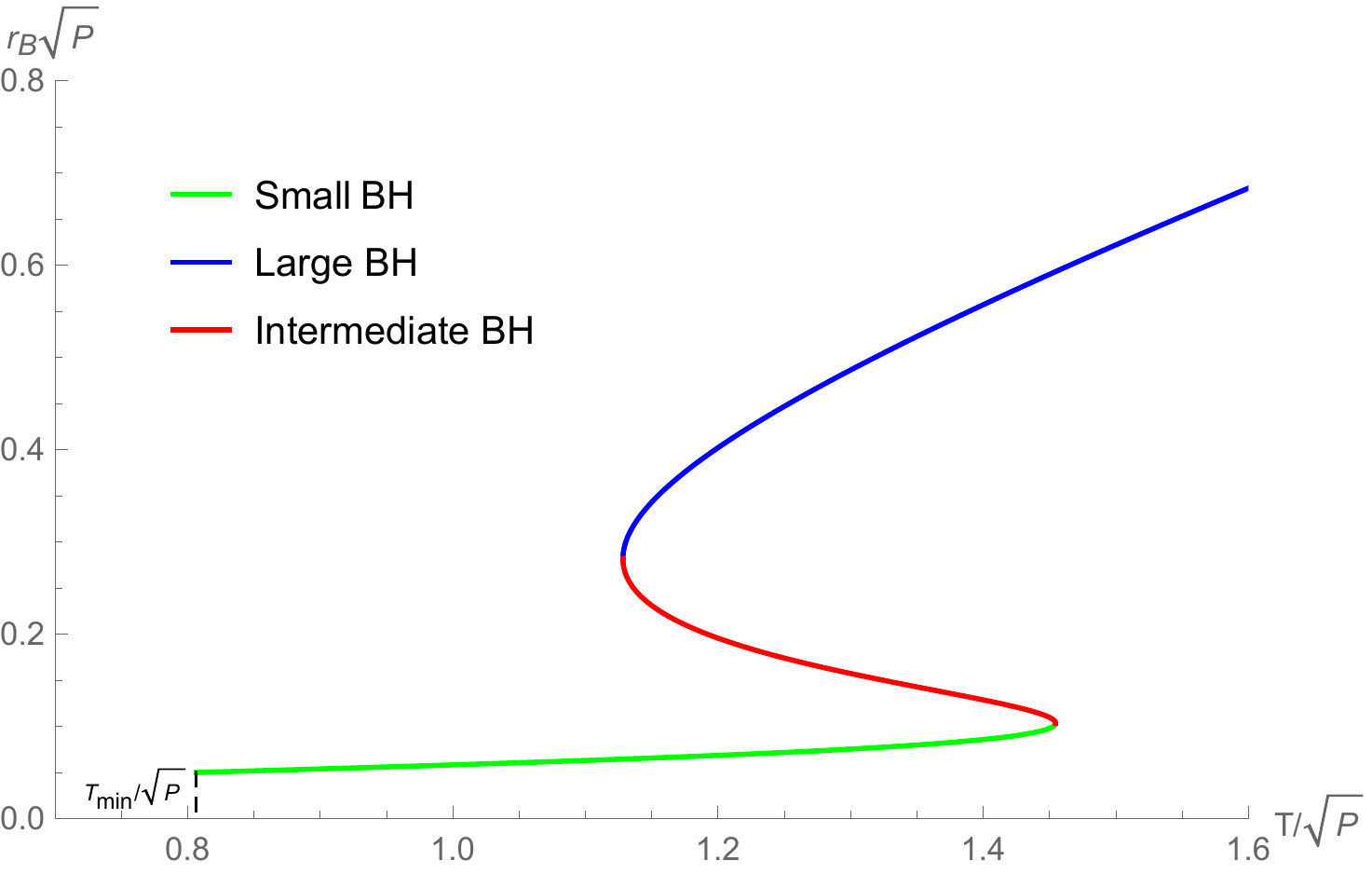}\label{fig:RNrS:b}
\end{center}
\caption{{\footnotesize Plots of $r_{h}/r_{B}\ $and $r_{B}\sqrt{P}$\ against
$T/\sqrt{P}$\ for a RN black hole in a cavity with $Q=0.05/\sqrt{P}<Q_{c}$, in
which three phases can coexist. The green, red and blue lines represent Small
BH, Intermediate BH and Large BH. Black hole solutions do not exist when
$T<T_{\min}$.}}%
\label{fig:RNrS}%
\end{figure}

Our strategy to discuss the system's phase structure and transitions under
constant pressure, temperature and charge will thus be the following: we start
from eqns. $\left(  \ref{eq:RNTQ}\right)  $ and $\left(  \ref{eq:RNP}\right)
$ to express $\tilde{T}$ as a function of $\tilde{r}_{B}$ and $\tilde{Q}$,
$\tilde{T}=\tilde{T}(\tilde{r}_{B},\tilde{Q})$. Similarly, using eqn. $\left(
\ref{eq:RNP}\right)  $, one can rewrite the Gibbs free energy and heat
capacity as $\tilde{G}=\tilde{G}(\tilde{r}_{B},\tilde{Q})$ and $\tilde{C}%
_{P}=\tilde{C}_{P}(\tilde{r}_{B},\tilde{Q})$, respectively. After $\tilde
{T}=\tilde{T}(\tilde{r}_{B},\tilde{Q})$ is solved for $\tilde{r}_{B}$ in terms
of $\tilde{T}$ and $\tilde{Q}$, we can express $\tilde{G}$ and $\tilde{C}_{P}$
with respect to $\tilde{T}$ and $\tilde{Q}$, $\tilde{G}=\tilde{G}(\tilde
{T},\tilde{Q})$ and $\tilde{C}_{P}=\tilde{C}_{P}(\tilde{T},\tilde{Q})$. A
critical point occurs at the inflection point of $\tilde{T}$ as a function of
$\tilde{r}_{B}$, where%
\begin{equation}
\frac{\partial\tilde{T}}{\partial\tilde{r}_{B}}=0\text{ and }\frac
{\partial^{2}\tilde{T}}{\partial\tilde{r}_{B}^{2}}=0.
\end{equation}
Solving the above equations gives quantities evaluated at the critical point%
\begin{equation}
(\tilde{r}_{Bc},\tilde{Q}_{c},\tilde{T}_{c},x_{c})\simeq\left(
0.222,0.107,1.100,0.911\right)  .
\end{equation}
The critical ratio, $P_{c}v_{c}/T_{c}\simeq0.405$, where $v\equiv2r_{B}$ is
the specific volume. Note that $P_{c}v_{c}/T_{c}=3/8$ for the van der Waals
fluid and a RN-AdS black hole.

\begin{figure}[tb]
\begin{center}
\includegraphics[width=0.48\textwidth]{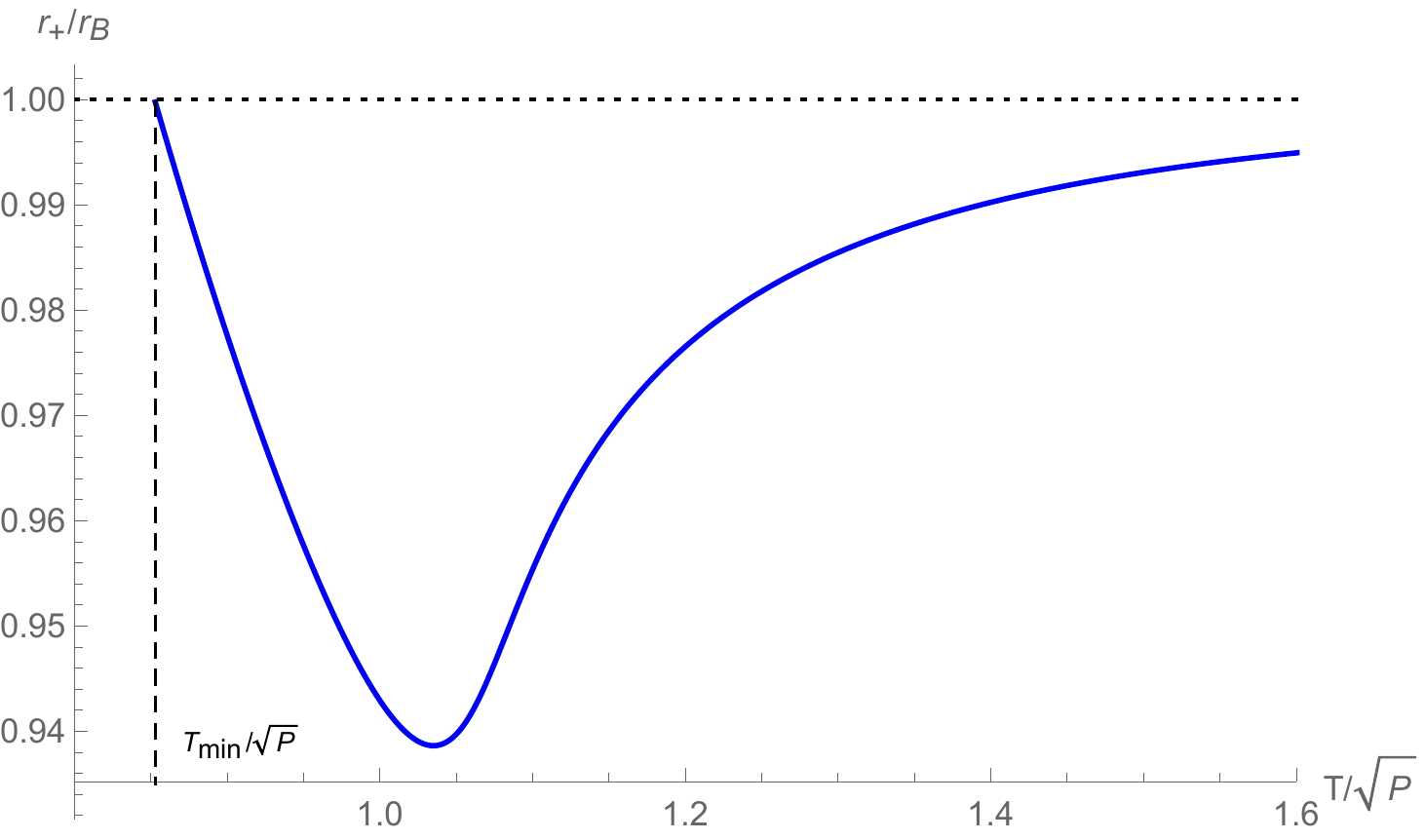}\label{fig:RNrL:a}
\includegraphics[width=0.48\textwidth]{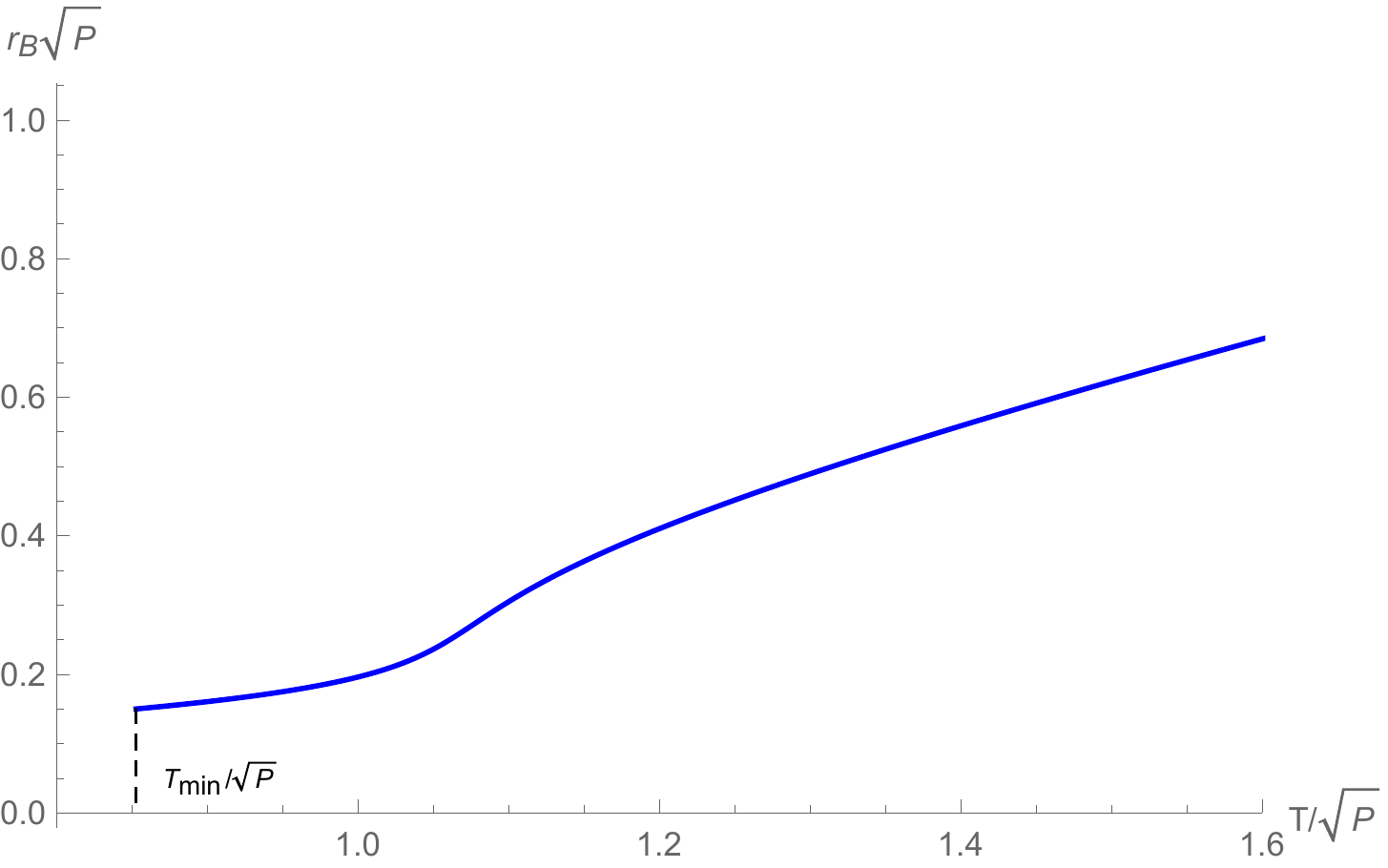}\label{fig:RNrL:b}
\end{center}
\caption{{\footnotesize Plots of $r_{h}/r_{B}\ $and $r_{B}\sqrt{P}$\ against
$T/\sqrt{P}$\ for a RN black hole in a cavity with $Q=0.15/\sqrt{P}>Q_{c}$.
There is only one phase, which does not exist when $T<T_{\min}$.}}%
\label{fig:RNrL}%
\end{figure}

With fixed $\tilde{T}$ and $\tilde{Q}$, the number of the phases,
corresponding to the branches of $\tilde{r}_{B}(\tilde{T},\tilde{Q})$, depends
on the value of $\tilde{Q}$. When $\tilde{Q}<\tilde{Q}_{c}$, three phases,
namely Small BH, Intermediate BH and Large BH, coexist for some range of $T$.
For $\tilde{Q}=0.05<\tilde{Q}_{c}$, we plot $x$ and $\tilde{r}_{B}$ against
$\tilde{T}$ in FIG. \ref{fig:RNrS}, where different colored lines represent
different phases. The first thing to note is that there exists a nonzero
minimum $\tilde{T}_{\min}$ for Small BH, which means that, unlike a RN-AdS
black hole, a RN black hole in a cavity can not become extremal if the
pressure is not zero. In other words, eqn. $\left(  \ref{eq:RNP}\right)  $
gives that an extremal RN black hole with $r_{+}=Q$ always has $P=0$. At
$\tilde{T}=\tilde{T}_{\min}$, it shows that $x=1$, which corresponds to the
black hole with the horizon merging with the wall of the cavity. As $\tilde
{T}$ increases, the horizon radius to cavity radius ratio increases toward $1$
for Large BH while decreases for Small BH and Intermediate BH. The cavity
enclosing Small BH and Large BH (Intermediate BH) expands (contracts) when
$\tilde{T}$ increases. Although it is not shown in the paper, we find that the
horizon radius has quite similar behavior to the cavity radius. In FIG.
\ref{fig:RNrL}, we display $x$ and $\tilde{r}_{B}$ against $\tilde{T}$ with
$\tilde{Q}=0.15>\tilde{Q}_{c}$, for which there exists only one phase. As one
increases $\tilde{T}$ from $\tilde{T}_{\min}$, the size of cavity keeps
growing. Meanwhile, the horizon radius to cavity radius ratio first decreases
from $1$ and then increases towards $1$.

\begin{figure}[tb]
\begin{center}
\includegraphics[width=0.48\textwidth]{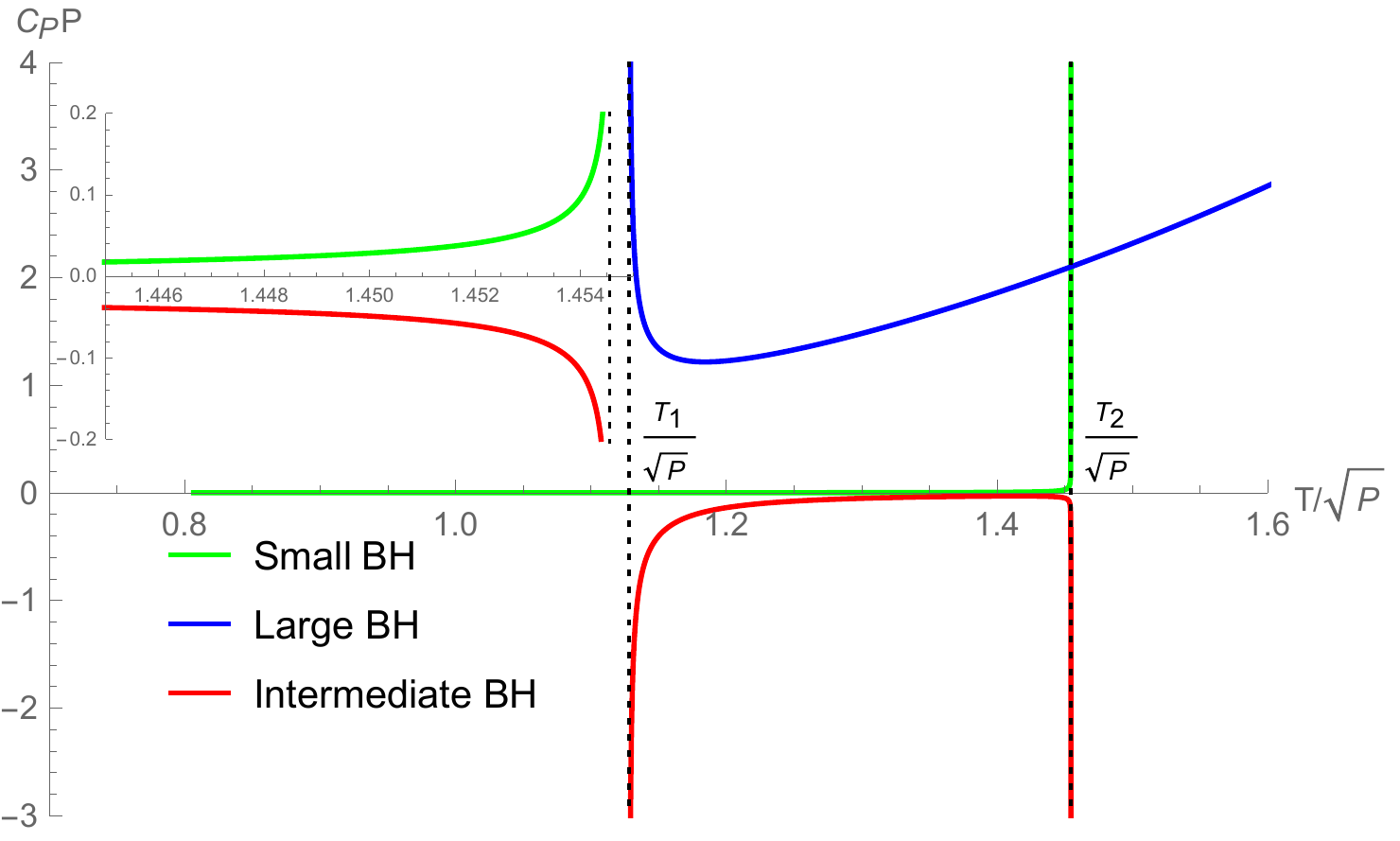}\label{fig:RNpS:a}
\includegraphics[width=0.48\textwidth]{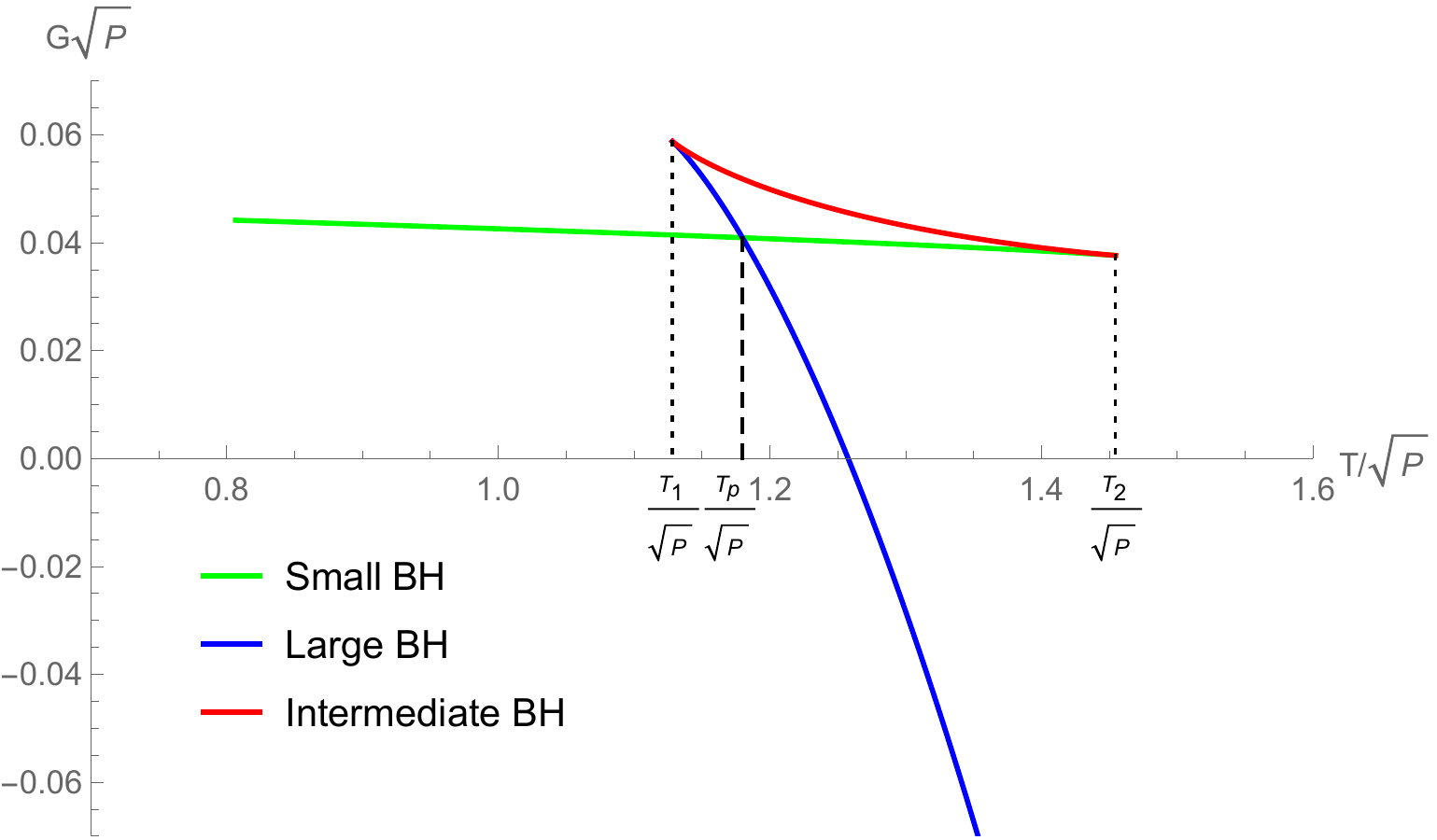}\label{fig:RNpS:b}
\end{center}
\caption{{\footnotesize Plots of $C_{p}P$ and $G\sqrt{P}$\ against $T/\sqrt
{P}$\ for a RN black hole in a cavity with $Q=0.05/\sqrt{P}<Q_{c}$.
\textbf{Left Panel}: Positive $C_{p}$ means that Large BH and Small BH are
thermally stable while negative $C_{p}$ gives Intermediate BH is thermally
unstable. \textbf{Right Panel}: As $T/\sqrt{P}$ increases from $T_{\min}%
/\sqrt{P}$, a first-order phase transition from Small BH to Large BH occurs at
$T=T_{p}$.}}%
\label{fig:RNpS}%
\end{figure}

\begin{figure}[tb]
\begin{center}
\includegraphics[width=0.48\textwidth]{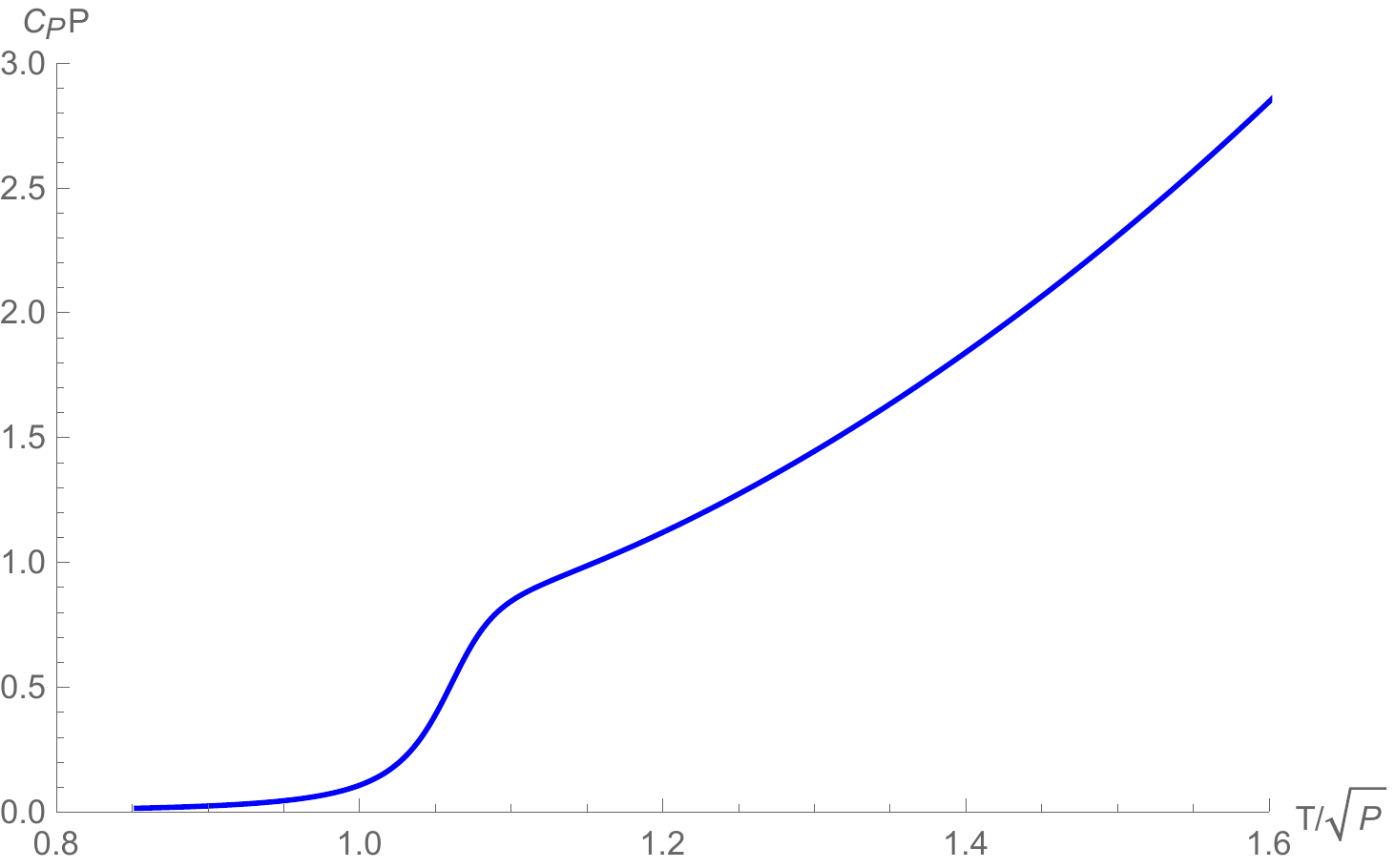}\label{fig:RNpL:a}
\includegraphics[width=0.48\textwidth]{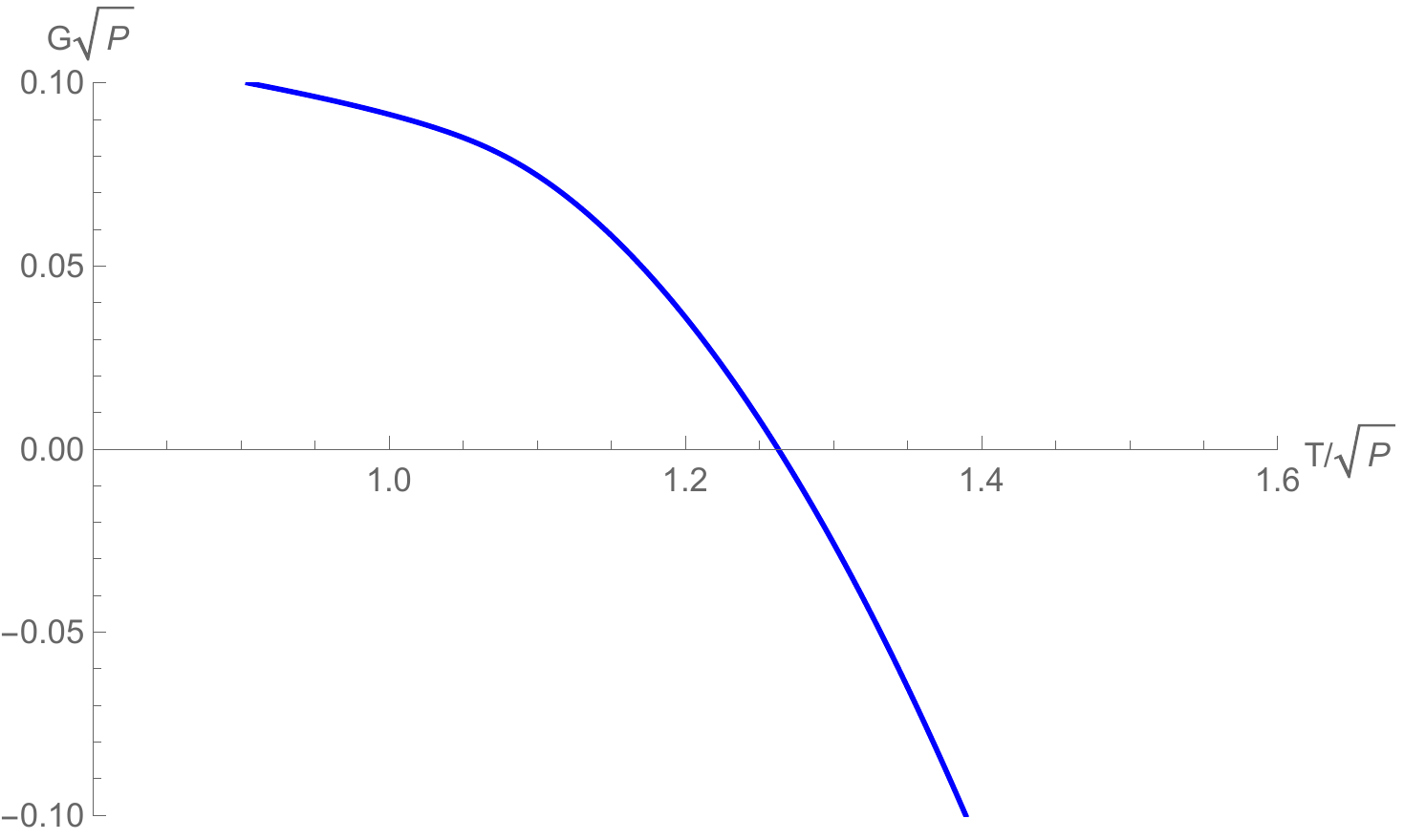}\label{fig:RNpL:b}
\end{center}
\caption{{\footnotesize Plots of $C_{p}P$ and $G\sqrt{P}$\ against $T/\sqrt
{P}$\ for a RN black hole in a cavity with $Q=0.15/\sqrt{P}>Q_{c}$.
\textbf{Left Panel}: The only phase has positive heat capacity and hence is
thermally stable. \textbf{Right Panel}: There is no phase transition.}}%
\label{fig:RNpL}%
\end{figure}

With $\tilde{r}_{B}(\tilde{T},\tilde{Q})$, we can obtain the heat capacity
$\tilde{C}_{P}(\tilde{T},\tilde{Q})$ and the Gibbs free energy $\tilde
{G}(\tilde{T},\tilde{Q})$ to discuss the stability of phases and phase
transitions. FIG. \ref{fig:RNpS} shows $\tilde{C}_{P}(\tilde{T},\tilde{Q})$
and $\tilde{G}(\tilde{T},\tilde{Q})$ in the left and right panels,
respectively, for a RN black hole in a cavity with $\tilde{Q}=0.05<\tilde
{Q}_{c}$. The heat capacity of Small BH and Large BH is positive, which means
that Small BH and Large BH are thermally stable. On the other hand,
Intermediate BH is a thermally unstable phase. Large BH, Small BH and
Intermediate BH coexist when $\tilde{T}_{1}<\tilde{T}<\tilde{T}_{2}$, and
there is a first-order phase transition between Small BH and Large BH
occurring at $\tilde{T}=\tilde{T}_{p}$ with $\tilde{T}_{1}<\tilde{T}%
_{p}<\tilde{T}_{2}$. The globally stable phase is Large BH for $T>T_{p}$ and
Small BH for $T<T_{p}$. For the system with $\tilde{Q}=0.15>\tilde{Q}_{c}$,
there is only one phase, which is always thermally stable (see FIG.
\ref{fig:RNpL}).

\begin{figure}[tb]
\begin{center}
\includegraphics[width=0.48\textwidth]{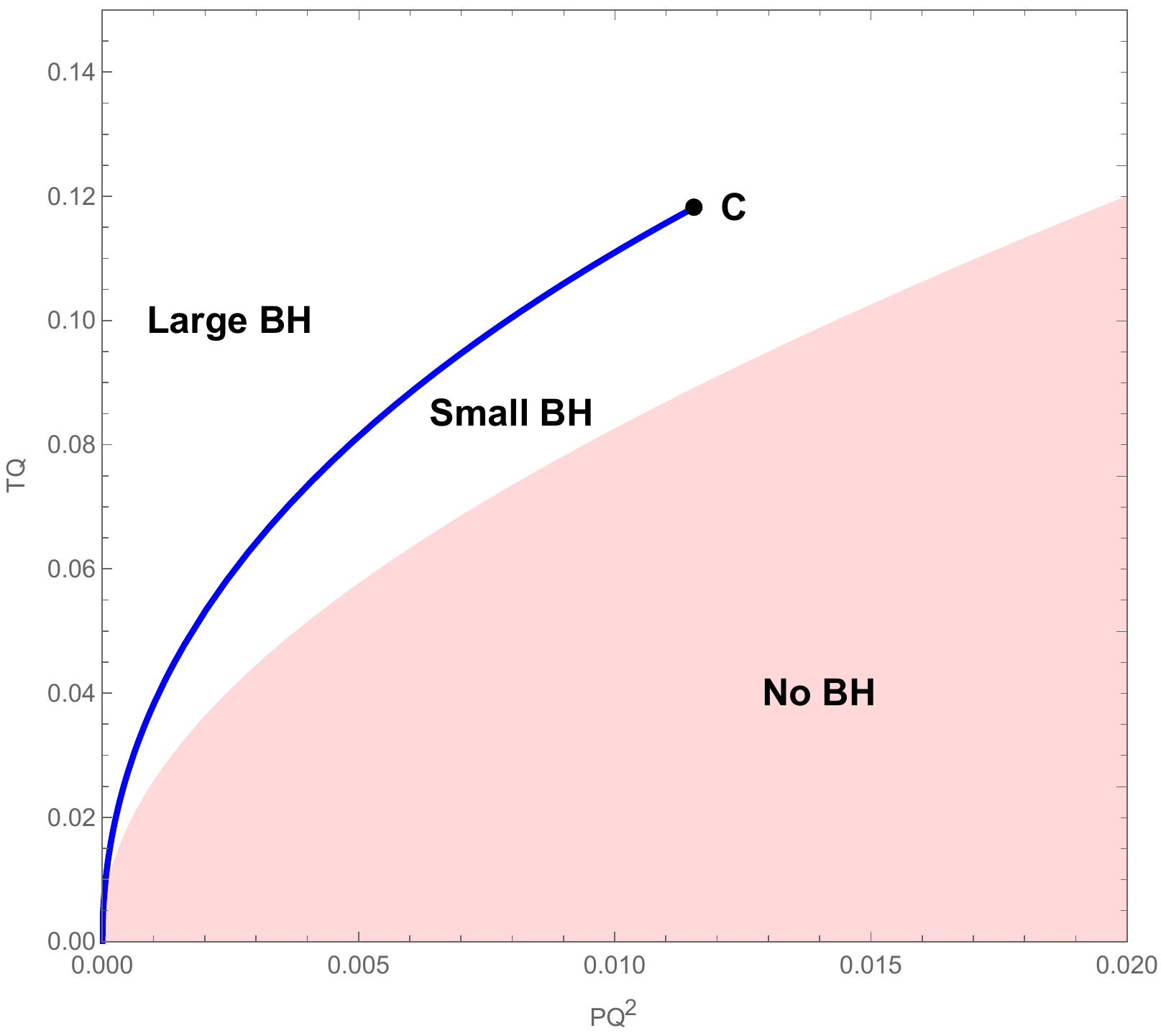}\label{fig:RNc:a}
\includegraphics[width=0.48\textwidth]{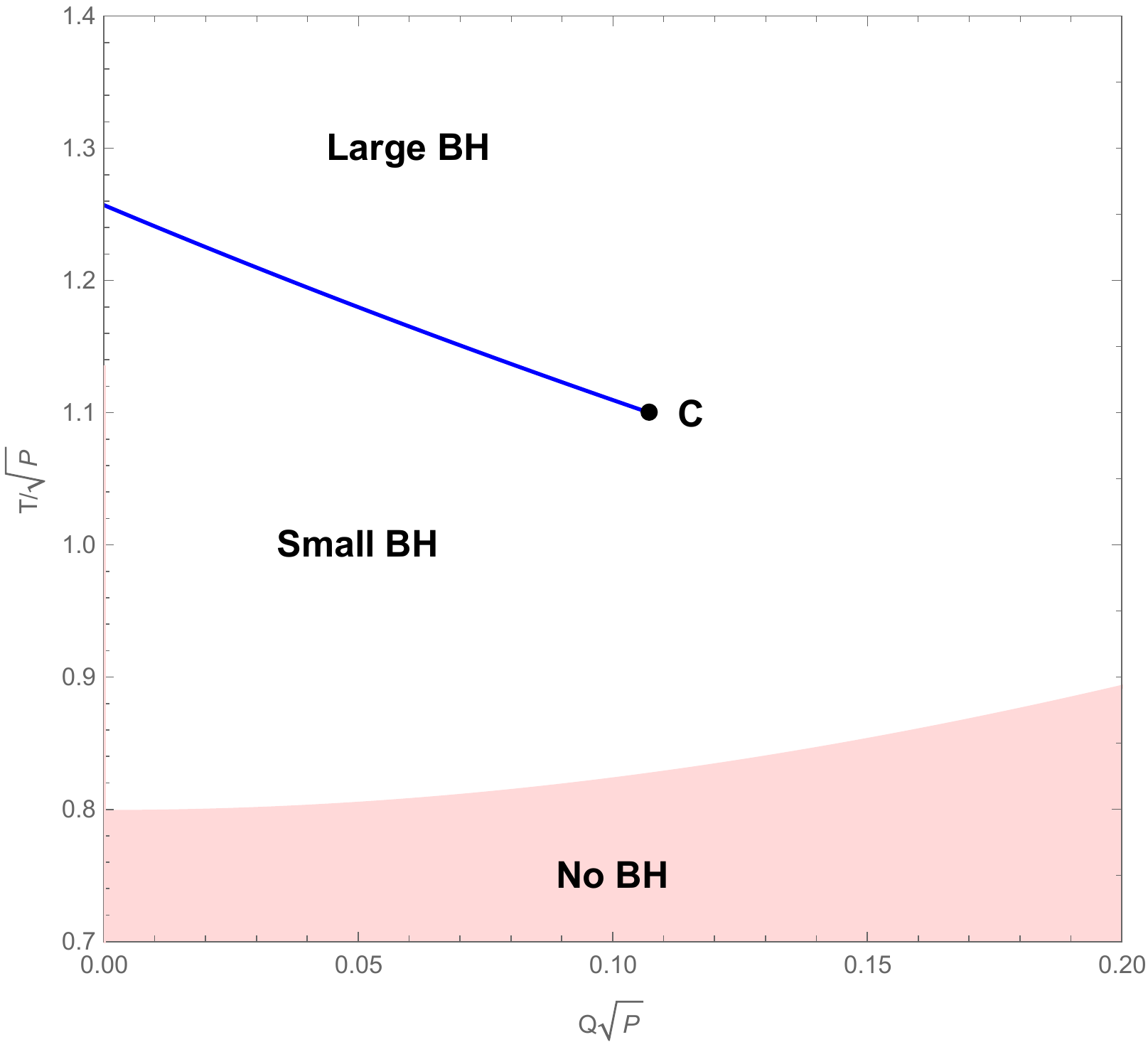}\label{fig:RNc:b}
\end{center}
\caption{{\footnotesize Phase diagrams of a RN-AdS black hole in a cavity in
the $PQ^{2}$-$TQ$ (\textbf{Left Panel}) and $Q\sqrt{P}$-$T/\sqrt{P}$
(\textbf{Right Panel}) planes. The first-order phase transition lines
separating Large BH and Small BH are displayed by blue lines and terminate at
the critical point, marked by black dots. The coexistence lines are of finite
length and reminiscent of the liquid/gas phase transition. No BH regions
correspond to $T<T_{\min}$, and hence no black hole solutions exist in light
red regions.}}%
\label{fig:RNc}%
\end{figure}

The left and right panels of FIG. \ref{fig:RNc} display the globally stable
phase of a RN black hole in a cavity in the $PQ^{2}$-$TQ$ and $\tilde{Q}%
$-$\tilde{T}$ planes, respectively. Blue lines represent Large BH/Small BH
first-order transition lines. These coexistence lines terminate at the
critical point, where a second-order phase transition occurs. These phase
diagrams show that the Large BH/Small BH phase transition is analogous to the
liquid/gas phase transition of the van der Waals fluid. Light red regions mark
no BH regions, where $T<T_{\min}$ and hence no black hole solutions exist.
Except the existence of no BH regions, these phase diagrams in the cavity case
are rather similar to those in the AdS case.

\begin{figure}[h]
\begin{center}
\includegraphics[width=0.45\textwidth]{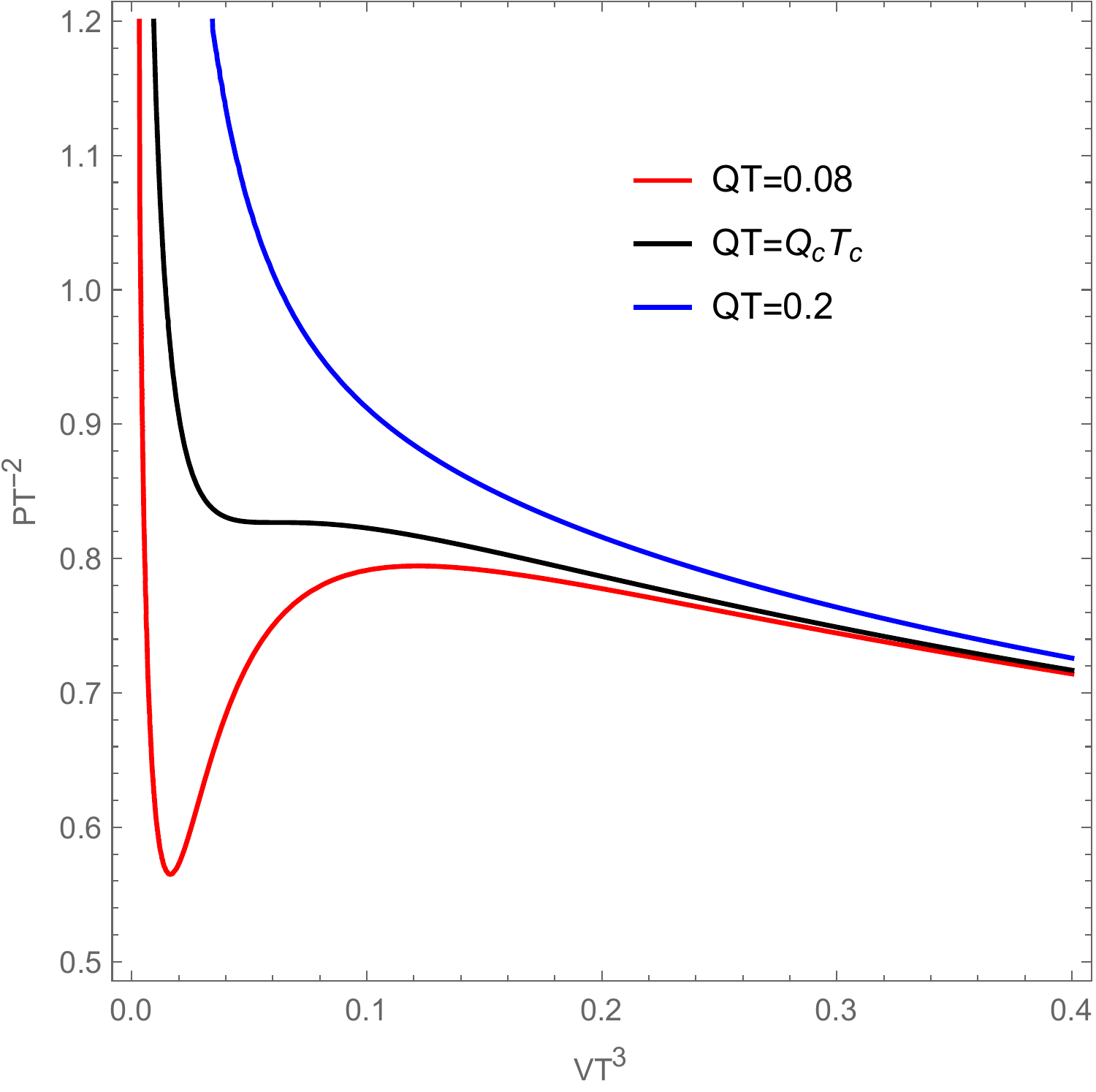}
\end{center}
\caption{{\footnotesize The equations of state for a RN black hole in a cavity
for different values of $QT$. The isotherms with fixed $Q$ bear a striking
resemblance to those of the Van der Waals fluid. The oscillating behavior
below $Q_{c}T_{c}$ is reminiscent of the liquid/gas phase transition of the
Van der Waals fluid.}}%
\label{fig:RNEoS}%
\end{figure}

Finally, we show further analogies with the van der Waals fluid in FIG.
\ref{fig:RNEoS}, where the equations of state with $QT$ ranging from below
(red) to above (blue) $Q_{c}T_{c}$ ($=\tilde{Q}_{c}\tilde{T}_{c}\simeq0.118$,
black) for a RN black in a cavity are plotted in the $VT^{3}$-$PT^{2}$ plane.
It is observed that these isotherms in the $VT^{3}$-$PT^{2}$ plane are
strikingly similar to those of van der Waals fluid. In fact, for a RN black in
a cavity with $T<\tilde{Q}_{c}\tilde{T}_{c}/Q$, the isotherm in the $VT^{3}%
$-$PT^{2}$ plane exhibits the oscillating part, in which a fixed $P$
corresponds to three $V$ solutions, namely Small BH, Intermediate BH and Large
BH. Such oscillation is reminiscent of the liquid/gas phase transition. Note
that the Maxwell's equal area law can determine the Large BH/Small BH phase
transition by defining a certain isobar to eliminate the oscillating part,
such that the areas above and below the isobar are equal. When $T>\tilde
{Q}_{c}\tilde{T}_{c}/Q$, the oscillating behavior of the isotherm disappears,
which is reminiscent of the ideal gas law.

\section{Discussion and Conclusion}

\label{Sec:DC}

In this paper, we extended the phase space of black holes enclosed in a cavity
to include $4\pi r_{B}^{3}/3$ as a thermodynamic volume, where $r_{B}$ is the
cavity radius. Consequently, the thermodynamic pressure $P$ is define by
$P=-\partial E/\partial V$, where $E$ is the thermal energy. Such extension is
largely motivated by the extended phase space of AdS black holes, in which the
cosmological constant\ is interpreted as a pressure. We showed that, in these
extended phase spaces, the thermodynamic behavior of black holes in a cavity
closely resembles that of the AdS counterparts, both exhibiting well-known
phenomena found in real-world systems.

Specifically, we first investigated phase structure of a Schwarzschild black
hole surrounded by a cavity in the extended phase space. We observed two
branches of black holes, Large BH and Small BH, above some temperature
$T_{\min}$, below which only the thermal flat spacetime exists. The Large BH
phase is thermally stable and has larger cavity and horizon radii. FIG.
\ref{fig:Schp} revealed that there is a first-order Hawking-Page-like phase
transition between the thermal spacetime and Large BH at $T_{p}>T_{\min}$.
Moreover, the coexistence line in the $P$-$T$ diagram is semi-infinite and
hence reminiscent of a solid/liquid phase transition. All these observations
indicate that the phase behavior of Schwarzschild black holes in a cavity and
Schwarzschild-AdS black holes is quite similar in the extended phase space.
However, we noticed in FIG. \ref{fig:SchEoS} that the isotherms of Small BH in
the $P$-$V$ diagram are significantly different in the two cases.

Considering a RN black hole enclosed by a cavity, we demonstrated that the
first-order phase transition between Large BH and Small BH is similar to the
van der Waals phase transition between liquid and gas, in that the Large
BH/Small BH coexistence lines in the $T$-$P$ and $T$-$Q$ diagrams (see FIG.
\ref{fig:RNc}) are of finite length and end at a second-order critical point.
FIG. \ref{fig:RNEoS} lent further evidence by showing that the isotherms of
the RN black hole in a cavity with fixed $Q$ in the $P$-$V$ diagram behave as
those of a van der Waals fluid system. Unlike RN-AdS black holes, RN black
holes in a cavity are not admissible in some $\left(  Q,P,T\right)
$-parameter region (e.g., RN black holes in a cavity with nonzero $P$ can not
become extremal). In spite of this difference, the thermodynamic properties of
RN black holes in a cavity and RN-AdS black holes are very much alike.

In existing studies, thermodynamic properties of black holes in a cavity were
investigated in the normal phase space, in which the cavity radius is fixed,
and found to closely resemble those of the AdS counterparts for various black
holes. In this paper, we extended the phase space to include the thermodynamic
volume and showed that analogies with the AdS counterparts still hold for
Schwarzschild and RN black holes in the extended phase space. On the other
hand, AdS black holes have been observed to possess a lot richer phase
structure and transitions, e.g., reentrant phase transitions and triple
points, in the extended phase space. It would be very interesting to explore
these phenomena in the context of black holes in a cavity and check whether
analogies with the AdS counterparts can go beyond Schwarzschild and RN black holes.

\begin{acknowledgments}
We are grateful to Shuxuan Ying and Zhipeng Zhang for useful discussions and
valuable comments. This work is supported in part by NSFC (Grant No. 11875196,
11375121, 11947225 and 11005016).
\end{acknowledgments}


\begin{thebibliography}{10}

\bibitem{Hawking:1971tu}
S.W. Hawking.
\newblock {Gravitational radiation from colliding black holes}.
\newblock {\em Phys. Rev. Lett.}, 26:1344--1346, 1971.
\newblock \href {https://doi.org/10.1103/PhysRevLett.26.1344}
  {\path{doi:10.1103/PhysRevLett.26.1344}}.

\bibitem{Bekenstein:1972tm}
Jacob~D Bekenstein.
\newblock Black holes and the second law.
\newblock {\em Lett. Nuovo Cim.}, 4(15):737--740, 1972.
\newblock \href {https://doi.org/10.1007/BF02757029}
  {\path{doi:10.1007/BF02757029}}.

\bibitem{Bekenstein:1973ur}
Jacob~D. Bekenstein.
\newblock {Black holes and entropy}.
\newblock {\em Phys. Rev. D}, 7:2333--2346, Apr 1973.
\newblock URL: \url{https://link.aps.org/doi/10.1103/PhysRevD.7.2333}, \href
  {https://doi.org/10.1103/PhysRevD.7.2333}
  {\path{doi:10.1103/PhysRevD.7.2333}}.

\bibitem{Hawking:1974rv}
S.W. Hawking.
\newblock {Black hole explosions}.
\newblock {\em Nature}, 248:30--31, 1974.
\newblock \href {https://doi.org/10.1038/248030a0}
  {\path{doi:10.1038/248030a0}}.

\bibitem{Hawking:1975iha}
S.W. Hawking.
\newblock {Particle Creation by Black Holes}.
\newblock In {\em {1st Oxford Conference on Quantum Gravity}}, pages 219--267,
  8 1975.

\bibitem{Bardeen:1973gs}
James~M. Bardeen, B.~Carter, and S.W. Hawking.
\newblock {The Four laws of black hole mechanics}.
\newblock {\em Commun. Math. Phys.}, 31:161--170, 1973.
\newblock \href {https://doi.org/10.1007/BF01645742}
  {\path{doi:10.1007/BF01645742}}.

\bibitem{Maldacena:1997re}
Juan~Martin Maldacena.
\newblock {The Large N limit of superconformal field theories and
  supergravity}.
\newblock {\em Int. J. Theor. Phys.}, 38:1113--1133, 1999.
\newblock \href {http://arxiv.org/abs/hep-th/9711200}
  {\path{arXiv:hep-th/9711200}}, \href
  {https://doi.org/10.1023/A:1026654312961}
  {\path{doi:10.1023/A:1026654312961}}.

\bibitem{Hawking:1982dh}
S.W. Hawking and Don~N. Page.
\newblock {Thermodynamics of Black Holes in anti-De Sitter Space}.
\newblock {\em Commun. Math. Phys.}, 87:577, 1983.
\newblock \href {https://doi.org/10.1007/BF01208266}
  {\path{doi:10.1007/BF01208266}}.

\bibitem{Witten:1998zw}
Edward Witten.
\newblock {Anti-de Sitter space, thermal phase transition, and confinement in
  gauge theories}.
\newblock {\em Adv. Theor. Math. Phys.}, 2:505--532, 1998.
\newblock \href {http://arxiv.org/abs/hep-th/9803131}
  {\path{arXiv:hep-th/9803131}}, \href
  {https://doi.org/10.4310/ATMP.1998.v2.n3.a3}
  {\path{doi:10.4310/ATMP.1998.v2.n3.a3}}.

\bibitem{Cvetic:1999ne}
Mirjam Cvetic and Steven~S. Gubser.
\newblock {Phases of R charged black holes, spinning branes and strongly
  coupled gauge theories}.
\newblock {\em JHEP}, 04:024, 1999.
\newblock \href {http://arxiv.org/abs/hep-th/9902195}
  {\path{arXiv:hep-th/9902195}}, \href
  {https://doi.org/10.1088/1126-6708/1999/04/024}
  {\path{doi:10.1088/1126-6708/1999/04/024}}.

\bibitem{Chamblin:1999tk}
Andrew Chamblin, Roberto Emparan, Clifford~V. Johnson, and Robert~C. Myers.
\newblock {Charged AdS black holes and catastrophic holography}.
\newblock {\em Phys. Rev. D}, 60:064018, Aug 1999.
\newblock URL: \url{https://link.aps.org/doi/10.1103/PhysRevD.60.064018}, \href
  {http://arxiv.org/abs/hep-th/9902170} {\path{arXiv:hep-th/9902170}}, \href
  {https://doi.org/10.1103/PhysRevD.60.064018}
  {\path{doi:10.1103/PhysRevD.60.064018}}.

\bibitem{Chamblin:1999hg}
Andrew Chamblin, Roberto Emparan, Clifford~V. Johnson, and Robert~C. Myers.
\newblock {Holography, thermodynamics and fluctuations of charged AdS black
  holes}.
\newblock {\em Phys. Rev. D}, 60:104026, Oct 1999.
\newblock URL: \url{https://link.aps.org/doi/10.1103/PhysRevD.60.104026}, \href
  {http://arxiv.org/abs/hep-th/9904197} {\path{arXiv:hep-th/9904197}}, \href
  {https://doi.org/10.1103/PhysRevD.60.104026}
  {\path{doi:10.1103/PhysRevD.60.104026}}.

\bibitem{Caldarelli:1999xj}
Marco~M. Caldarelli, Guido Cognola, and Dietmar Klemm.
\newblock {Thermodynamics of Kerr-Newman-AdS black holes and conformal field
  theories}.
\newblock {\em Class. Quant. Grav.}, 17:399--420, 2000.
\newblock \href {http://arxiv.org/abs/hep-th/9908022}
  {\path{arXiv:hep-th/9908022}}, \href
  {https://doi.org/10.1088/0264-9381/17/2/310}
  {\path{doi:10.1088/0264-9381/17/2/310}}.

\bibitem{Cai:2001dz}
Rong-Gen Cai.
\newblock {Gauss-Bonnet black holes in AdS spaces}.
\newblock {\em Phys. Rev. D}, 65:084014, 2002.
\newblock \href {http://arxiv.org/abs/hep-th/0109133}
  {\path{arXiv:hep-th/0109133}}, \href
  {https://doi.org/10.1103/PhysRevD.65.084014}
  {\path{doi:10.1103/PhysRevD.65.084014}}.

\bibitem{Cvetic:2001bk}
Mirjam Cvetic, Shin'ichi Nojiri, and Sergei~D. Odintsov.
\newblock {Black hole thermodynamics and negative entropy in de Sitter and
  anti-de Sitter Einstein-Gauss-Bonnet gravity}.
\newblock {\em Nucl. Phys. B}, 628:295--330, 2002.
\newblock \href {http://arxiv.org/abs/hep-th/0112045}
  {\path{arXiv:hep-th/0112045}}, \href
  {https://doi.org/10.1016/S0550-3213(02)00075-5}
  {\path{doi:10.1016/S0550-3213(02)00075-5}}.

\bibitem{Nojiri:2001aj}
Shin'ichi Nojiri and Sergei~D. Odintsov.
\newblock {Anti-de Sitter black hole thermodynamics in higher derivative
  gravity and new confining deconfining phases in dual CFT}.
\newblock {\em Phys. Lett. B}, 521:87--95, 2001.
\newblock [Erratum: Phys.Lett.B 542, 301 (2002)].
\newblock \href {http://arxiv.org/abs/hep-th/0109122}
  {\path{arXiv:hep-th/0109122}}, \href
  {https://doi.org/10.1016/S0370-2693(01)01186-8}
  {\path{doi:10.1016/S0370-2693(01)01186-8}}.

\bibitem{Peca:1998cs}
Claudia~S. Peca and Jose~P.S. Lemos.
\newblock {Thermodynamics of Reissner-Nordstrom anti-de Sitter black holes in
  the grand canonical ensemble}.
\newblock {\em Phys. Rev. D}, 59:124007, 1999.
\newblock \href {http://arxiv.org/abs/gr-qc/9805004}
  {\path{arXiv:gr-qc/9805004}}, \href
  {https://doi.org/10.1103/PhysRevD.59.124007}
  {\path{doi:10.1103/PhysRevD.59.124007}}.

\bibitem{Dolan:2011xt}
Brian~P. Dolan.
\newblock {Pressure and volume in the first law of black hole thermodynamics}.
\newblock {\em Class. Quant. Grav.}, 28:235017, 2011.
\newblock \href {http://arxiv.org/abs/1106.6260} {\path{arXiv:1106.6260}},
  \href {https://doi.org/10.1088/0264-9381/28/23/235017}
  {\path{doi:10.1088/0264-9381/28/23/235017}}.

\bibitem{Kubiznak:2012wp}
David Kubiznak and Robert~B. Mann.
\newblock {P-V criticality of charged AdS black holes}.
\newblock {\em JHEP}, 07:033, 2012.
\newblock \href {http://arxiv.org/abs/1205.0559} {\path{arXiv:1205.0559}},
  \href {https://doi.org/10.1007/JHEP07(2012)033}
  {\path{doi:10.1007/JHEP07(2012)033}}.

\bibitem{Kastor:2009wy}
David Kastor, Sourya Ray, and Jennie Traschen.
\newblock {Enthalpy and the Mechanics of AdS Black Holes}.
\newblock {\em Class. Quant. Grav.}, 26:195011, 2009.
\newblock \href {http://arxiv.org/abs/0904.2765} {\path{arXiv:0904.2765}},
  \href {https://doi.org/10.1088/0264-9381/26/19/195011}
  {\path{doi:10.1088/0264-9381/26/19/195011}}.

\bibitem{Wei:2012ui}
Shao-Wen Wei and Yu-Xiao Liu.
\newblock {Critical phenomena and thermodynamic geometry of charged
  Gauss-Bonnet AdS black holes}.
\newblock {\em Phys. Rev. D}, 87(4):044014, Feb 2013.
\newblock URL: \url{https://link.aps.org/doi/10.1103/PhysRevD.87.044014}, \href
  {http://arxiv.org/abs/1209.1707} {\path{arXiv:1209.1707}}, \href
  {https://doi.org/10.1103/PhysRevD.87.044014}
  {\path{doi:10.1103/PhysRevD.87.044014}}.

\bibitem{Gunasekaran:2012dq}
Sharmila Gunasekaran, Robert~B. Mann, and David Kubiznak.
\newblock {Extended phase space thermodynamics for charged and rotating black
  holes and Born-Infeld vacuum polarization}.
\newblock {\em JHEP}, 11:110, 2012.
\newblock \href {http://arxiv.org/abs/1208.6251} {\path{arXiv:1208.6251}},
  \href {https://doi.org/10.1007/JHEP11(2012)110}
  {\path{doi:10.1007/JHEP11(2012)110}}.

\bibitem{Cai:2013qga}
Rong-Gen Cai, Li-Ming Cao, Li~Li, and Run-Qiu Yang.
\newblock {P-V criticality in the extended phase space of Gauss-Bonnet black
  holes in AdS space}.
\newblock {\em JHEP}, 09:005, 2013.
\newblock \href {http://arxiv.org/abs/1306.6233} {\path{arXiv:1306.6233}},
  \href {https://doi.org/10.1007/JHEP09(2013)005}
  {\path{doi:10.1007/JHEP09(2013)005}}.

\bibitem{Xu:2014kwa}
Wei Xu and Liu Zhao.
\newblock {Critical phenomena of static charged AdS black holes in conformal
  gravity}.
\newblock {\em Phys. Lett. B}, 736:214--220, 2014.
\newblock \href {http://arxiv.org/abs/1405.7665} {\path{arXiv:1405.7665}},
  \href {https://doi.org/10.1016/j.physletb.2014.07.019}
  {\path{doi:10.1016/j.physletb.2014.07.019}}.

\bibitem{Frassino:2014pha}
Antonia~M. Frassino, David Kubiznak, Robert~B. Mann, and Fil Simovic.
\newblock {Multiple Reentrant Phase Transitions and Triple Points in Lovelock
  Thermodynamics}.
\newblock {\em JHEP}, 09:080, 2014.
\newblock \href {http://arxiv.org/abs/1406.7015} {\path{arXiv:1406.7015}},
  \href {https://doi.org/10.1007/JHEP09(2014)080}
  {\path{doi:10.1007/JHEP09(2014)080}}.

\bibitem{Dehghani:2014caa}
M.~H. Dehghani, S.~Kamrani, and A.~Sheykhi.
\newblock {$P-V$ criticality of charged dilatonic black holes}.
\newblock {\em Phys. Rev. D}, 90(10):104020, 2014.
\newblock \href {http://arxiv.org/abs/1505.02386} {\path{arXiv:1505.02386}},
  \href {https://doi.org/10.1103/PhysRevD.90.104020}
  {\path{doi:10.1103/PhysRevD.90.104020}}.

\bibitem{Hennigar:2015esa}
Robie~A. Hennigar, Wilson~G. Brenna, and Robert~B. Mann.
\newblock {$P-v$ criticality in quasitopological gravity}.
\newblock {\em JHEP}, 07:077, 2015.
\newblock \href {http://arxiv.org/abs/1505.05517} {\path{arXiv:1505.05517}},
  \href {https://doi.org/10.1007/JHEP07(2015)077}
  {\path{doi:10.1007/JHEP07(2015)077}}.

\bibitem{Caceres:2015vsa}
Elena Caceres, Phuc~H. Nguyen, and Juan~F. Pedraza.
\newblock {Holographic entanglement entropy and the extended phase structure of
  STU black holes}.
\newblock {\em JHEP}, 09:184, 2015.
\newblock \href {http://arxiv.org/abs/1507.06069} {\path{arXiv:1507.06069}},
  \href {https://doi.org/10.1007/JHEP09(2015)184}
  {\path{doi:10.1007/JHEP09(2015)184}}.

\bibitem{Hendi:2016yof}
Seyed~Hossein Hendi, Gu-Qiang Li, Jie-Xiong Mo, Shahram Panahiyan, and Behzad
  Eslam~Panah.
\newblock {New perspective for black hole thermodynamics in
  Gauss--Bonnet--Born--Infeld massive gravity}.
\newblock {\em Eur. Phys. J. C}, 76(10):571, 2016.
\newblock \href {http://arxiv.org/abs/1608.03148} {\path{arXiv:1608.03148}},
  \href {https://doi.org/10.1140/epjc/s10052-016-4410-4}
  {\path{doi:10.1140/epjc/s10052-016-4410-4}}.

\bibitem{Hendi:2017fxp}
S.H. Hendi, R.B. Mann, S.~Panahiyan, and B.~Eslam~Panah.
\newblock {Van der Waals like behavior of topological AdS black holes in
  massive gravity}.
\newblock {\em Phys. Rev. D}, 95(2):021501, 2017.
\newblock \href {http://arxiv.org/abs/1702.00432} {\path{arXiv:1702.00432}},
  \href {https://doi.org/10.1103/PhysRevD.95.021501}
  {\path{doi:10.1103/PhysRevD.95.021501}}.

\bibitem{Lemos:2018cfd}
José~P.S. Lemos and Oleg~B. Zaslavskii.
\newblock {Black hole thermodynamics with the cosmological constant as
  independent variable: Bridge between the enthalpy and the Euclidean path
  integral approaches}.
\newblock {\em Phys. Lett. B}, 786:296--299, 2018.
\newblock \href {http://arxiv.org/abs/1806.07910} {\path{arXiv:1806.07910}},
  \href {https://doi.org/10.1016/j.physletb.2018.08.075}
  {\path{doi:10.1016/j.physletb.2018.08.075}}.

\bibitem{Pedraza:2018eey}
Juan~F. Pedraza, Watse Sybesma, and Manus~R. Visser.
\newblock {Hyperscaling violating black holes with spherical and hyperbolic
  horizons}.
\newblock {\em Class. Quant. Grav.}, 36(5):054002, 2019.
\newblock \href {http://arxiv.org/abs/1807.09770} {\path{arXiv:1807.09770}},
  \href {https://doi.org/10.1088/1361-6382/ab0094}
  {\path{doi:10.1088/1361-6382/ab0094}}.

\bibitem{Wang:2018xdz}
Peng Wang, Houwen Wu, and Haitang Yang.
\newblock {Thermodynamics and Phase Transitions of Nonlinear Electrodynamics
  Black Holes in an Extended Phase Space}.
\newblock {\em JCAP}, 04(04):052, 2019.
\newblock \href {http://arxiv.org/abs/1808.04506} {\path{arXiv:1808.04506}},
  \href {https://doi.org/10.1088/1475-7516/2019/04/052}
  {\path{doi:10.1088/1475-7516/2019/04/052}}.

\bibitem{Wei:2020poh}
Shao-Wen Wei and Yu-Xiao Liu.
\newblock {Extended thermodynamics and microstructures of four-dimensional
  charged Gauss-Bonnet black hole in AdS space}.
\newblock {\em Phys. Rev. D}, 101(10):104018, 2020.
\newblock \href {http://arxiv.org/abs/2003.14275} {\path{arXiv:2003.14275}},
  \href {https://doi.org/10.1103/PhysRevD.101.104018}
  {\path{doi:10.1103/PhysRevD.101.104018}}.

\bibitem{Kubiznak:2016qmn}
David Kubiznak, Robert~B. Mann, and Mae Teo.
\newblock {Black hole chemistry: thermodynamics with Lambda}.
\newblock {\em Class. Quant. Grav.}, 34(6):063001, 2017.
\newblock \href {http://arxiv.org/abs/1608.06147} {\path{arXiv:1608.06147}},
  \href {https://doi.org/10.1088/1361-6382/aa5c69}
  {\path{doi:10.1088/1361-6382/aa5c69}}.

\bibitem{Kubiznak:2014zwa}
David Kubiznak and Robert~B. Mann.
\newblock {Black hole chemistry}.
\newblock {\em Can. J. Phys.}, 93(9):999--1002, 2015.
\newblock \href {http://arxiv.org/abs/1404.2126} {\path{arXiv:1404.2126}},
  \href {https://doi.org/10.1139/cjp-2014-0465}
  {\path{doi:10.1139/cjp-2014-0465}}.

\bibitem{York:1986it}
Jr. York, James~W.
\newblock {Black hole thermodynamics and the Euclidean Einstein action}.
\newblock {\em Phys. Rev. D}, 33:2092--2099, Apr 1986.
\newblock URL: \url{https://link.aps.org/doi/10.1103/PhysRevD.33.2092}, \href
  {https://doi.org/10.1103/PhysRevD.33.2092}
  {\path{doi:10.1103/PhysRevD.33.2092}}.

\bibitem{Braden:1990hw}
Harry~W. Braden, J.~David Brown, Bernard~F. Whiting, and James~W. York.
\newblock Charged black hole in a grand canonical ensemble.
\newblock {\em Phys. Rev. D}, 42:3376--3385, Nov 1990.
\newblock URL: \url{https://link.aps.org/doi/10.1103/PhysRevD.42.3376}, \href
  {https://doi.org/10.1103/PhysRevD.42.3376}
  {\path{doi:10.1103/PhysRevD.42.3376}}.

\bibitem{Carlip:2003ne}
Steven Carlip and S.~Vaidya.
\newblock {Phase transitions and critical behavior for charged black holes}.
\newblock {\em Class. Quant. Grav.}, 20:3827--3838, 2003.
\newblock \href {http://arxiv.org/abs/gr-qc/0306054}
  {\path{arXiv:gr-qc/0306054}}, \href
  {https://doi.org/10.1088/0264-9381/20/16/319}
  {\path{doi:10.1088/0264-9381/20/16/319}}.

\bibitem{Lundgren:2006kt}
Andrew~P. Lundgren.
\newblock {Charged black hole in a canonical ensemble}.
\newblock {\em Phys. Rev. D}, 77:044014, 2008.
\newblock \href {http://arxiv.org/abs/gr-qc/0612119}
  {\path{arXiv:gr-qc/0612119}}, \href
  {https://doi.org/10.1103/PhysRevD.77.044014}
  {\path{doi:10.1103/PhysRevD.77.044014}}.

\bibitem{Lu:2010xt}
J.X. Lu, Shibaji Roy, and Zhiguang Xiao.
\newblock {Phase transitions and critical behavior of black branes in canonical
  ensemble}.
\newblock {\em JHEP}, 01:133, 2011.
\newblock \href {http://arxiv.org/abs/1010.2068} {\path{arXiv:1010.2068}},
  \href {https://doi.org/10.1007/JHEP01(2011)133}
  {\path{doi:10.1007/JHEP01(2011)133}}.

\bibitem{Wu:2011yu}
Chao Wu, Zhiguang Xiao, and Jianfei Xu.
\newblock {Bubbles and Black Branes in Grand Canonical Ensemble}.
\newblock {\em Phys. Rev. D}, 85:044009, 2012.
\newblock \href {http://arxiv.org/abs/1108.1347} {\path{arXiv:1108.1347}},
  \href {https://doi.org/10.1103/PhysRevD.85.044009}
  {\path{doi:10.1103/PhysRevD.85.044009}}.

\bibitem{Lu:2012rm}
J.X. Lu, Ran Wei, and Jianfei Xu.
\newblock {The phase structure of black D1/D5 (F/NS5) system in canonical
  ensemble}.
\newblock {\em JHEP}, 12:012, 2012.
\newblock \href {http://arxiv.org/abs/1210.0708} {\path{arXiv:1210.0708}},
  \href {https://doi.org/10.1007/JHEP12(2012)012}
  {\path{doi:10.1007/JHEP12(2012)012}}.

\bibitem{Lu:2013nt}
J.X. Lu and Ran Wei.
\newblock {Modulating the phase structure of black D6 branes in canonical
  ensemble}.
\newblock {\em JHEP}, 04:100, 2013.
\newblock \href {http://arxiv.org/abs/1301.1780} {\path{arXiv:1301.1780}},
  \href {https://doi.org/10.1007/JHEP04(2013)100}
  {\path{doi:10.1007/JHEP04(2013)100}}.

\bibitem{Zhou:2015yxa}
Da~Zhou and Zhiguang Xiao.
\newblock {Phase structures of the black Dp-D(p + 4)-brane system in various
  ensembles I: thermal stability}.
\newblock {\em JHEP}, 07:134, 2015.
\newblock \href {http://arxiv.org/abs/1502.00261} {\path{arXiv:1502.00261}},
  \href {https://doi.org/10.1007/JHEP07(2015)134}
  {\path{doi:10.1007/JHEP07(2015)134}}.

\bibitem{Xiao:2015bha}
Zhiguang Xiao and Da~Zhou.
\newblock {Phase structures of the black D$p$-D$(p + 4)$-brane system in
  various ensembles II: electrical and thermodynamic stability}.
\newblock {\em JHEP}, 09:028, 2015.
\newblock \href {http://arxiv.org/abs/1507.02088} {\path{arXiv:1507.02088}},
  \href {https://doi.org/10.1007/JHEP09(2015)028}
  {\path{doi:10.1007/JHEP09(2015)028}}.

\bibitem{Sanchis-Gual:2015lje}
Nicolas Sanchis-Gual, Juan~Carlos Degollado, Pedro~J. Montero, José~A. Font,
  and Carlos Herdeiro.
\newblock {Explosion and Final State of an Unstable Reissner-Nordström Black
  Hole}.
\newblock {\em Phys. Rev. Lett.}, 116(14):141101, 2016.
\newblock \href {http://arxiv.org/abs/1512.05358} {\path{arXiv:1512.05358}},
  \href {https://doi.org/10.1103/PhysRevLett.116.141101}
  {\path{doi:10.1103/PhysRevLett.116.141101}}.

\bibitem{Sanchis-Gual:2016tcm}
Nicolas Sanchis-Gual, Juan~Carlos Degollado, Carlos Herdeiro, José~A. Font,
  and Pedro~J. Montero.
\newblock {Dynamical formation of a Reissner-Nordström black hole with scalar
  hair in a cavity}.
\newblock {\em Phys. Rev. D}, 94(4):044061, 2016.
\newblock \href {http://arxiv.org/abs/1607.06304} {\path{arXiv:1607.06304}},
  \href {https://doi.org/10.1103/PhysRevD.94.044061}
  {\path{doi:10.1103/PhysRevD.94.044061}}.

\bibitem{Basu:2016srp}
Pallab Basu, Chethan Krishnan, and P.~N. Bala~Subramanian.
\newblock {Hairy Black Holes in a Box}.
\newblock {\em JHEP}, 11:041, 2016.
\newblock \href {http://arxiv.org/abs/1609.01208} {\path{arXiv:1609.01208}},
  \href {https://doi.org/10.1007/JHEP11(2016)041}
  {\path{doi:10.1007/JHEP11(2016)041}}.

\bibitem{Peng:2017gss}
Yan Peng, Bin Wang, and Yunqi Liu.
\newblock {On the thermodynamics of the black hole and hairy black hole
  transitions in the asymptotically flat spacetime with a box}.
\newblock {\em Eur. Phys. J. C}, 78(3):176, 2018.
\newblock \href {http://arxiv.org/abs/1708.01411} {\path{arXiv:1708.01411}},
  \href {https://doi.org/10.1140/epjc/s10052-018-5652-0}
  {\path{doi:10.1140/epjc/s10052-018-5652-0}}.

\bibitem{Peng:2017squ}
Yan Peng.
\newblock {Studies of a general flat space/boson star transition model in a box
  through a language similar to holographic superconductors}.
\newblock {\em JHEP}, 07:042, 2017.
\newblock \href {http://arxiv.org/abs/1705.08694} {\path{arXiv:1705.08694}},
  \href {https://doi.org/10.1007/JHEP07(2017)042}
  {\path{doi:10.1007/JHEP07(2017)042}}.

\bibitem{Peng:2018abh}
Yan Peng.
\newblock {Scalar field configurations supported by charged compact reflecting
  stars in a curved spacetime}.
\newblock {\em Phys. Lett. B}, 780:144--148, 2018.
\newblock \href {http://arxiv.org/abs/1801.02495} {\path{arXiv:1801.02495}},
  \href {https://doi.org/10.1016/j.physletb.2018.02.068}
  {\path{doi:10.1016/j.physletb.2018.02.068}}.

\bibitem{Kiczek:2019qbk}
Bartlomiej Kiczek and Marek Rogatko.
\newblock {Ultra-compact spherically symmetric dark matter charged star
  objects}.
\newblock {\em JCAP}, 09:049, 2019.
\newblock \href {http://arxiv.org/abs/1904.07232} {\path{arXiv:1904.07232}},
  \href {https://doi.org/10.1088/1475-7516/2019/09/049}
  {\path{doi:10.1088/1475-7516/2019/09/049}}.

\bibitem{Kiczek:2020gyd}
Bartlomiej Kiczek and Marek Rogatko.
\newblock {Influence of dark matter on black hole scalar hair}.
\newblock {\em Phys. Rev. D}, 101(8):084035, 2020.
\newblock \href {http://arxiv.org/abs/2004.06617} {\path{arXiv:2004.06617}},
  \href {https://doi.org/10.1103/PhysRevD.101.084035}
  {\path{doi:10.1103/PhysRevD.101.084035}}.

\bibitem{Wang:2019urm}
Peng Wang, Haitang Yang, and Shuxuan Ying.
\newblock {Thermodynamics and phase transition of a Gauss-Bonnet black hole in
  a cavity}.
\newblock {\em Phys. Rev. D}, 101(6):064045, 2020.
\newblock \href {http://arxiv.org/abs/1909.01275} {\path{arXiv:1909.01275}},
  \href {https://doi.org/10.1103/PhysRevD.101.064045}
  {\path{doi:10.1103/PhysRevD.101.064045}}.

\bibitem{Wang:2019kxp}
Peng Wang, Houwen Wu, and Haitang Yang.
\newblock {Thermodynamics and Phase Transition of a Nonlinear Electrodynamics
  Black Hole in a Cavity}.
\newblock {\em JHEP}, 07:002, 2019.
\newblock \href {http://arxiv.org/abs/1901.06216} {\path{arXiv:1901.06216}},
  \href {https://doi.org/10.1007/JHEP07(2019)002}
  {\path{doi:10.1007/JHEP07(2019)002}}.

\bibitem{Liang:2019dni}
Kangkai Liang, Peng Wang, Houwen Wu, and Mingtao Yang.
\newblock {Phase structures and transitions of Born--Infeld black holes in a
  grand canonical ensemble}.
\newblock {\em Eur. Phys. J. C}, 80(3):187, 2020.
\newblock \href {http://arxiv.org/abs/1907.00799} {\path{arXiv:1907.00799}},
  \href {https://doi.org/10.1140/epjc/s10052-020-7750-z}
  {\path{doi:10.1140/epjc/s10052-020-7750-z}}.

\bibitem{Wang:2019cax}
Peng Wang, Houwen Wu, and Haitang Yang.
\newblock {Thermodynamic Geometry of AdS Black Holes and Black Holes in a
  Cavity}.
\newblock {\em Eur. Phys. J. C}, 80(3):216, 2020.
\newblock \href {http://arxiv.org/abs/1910.07874} {\path{arXiv:1910.07874}},
  \href {https://doi.org/10.1140/epjc/s10052-020-7776-2}
  {\path{doi:10.1140/epjc/s10052-020-7776-2}}.

\bibitem{Wang:2020osg}
Peng Wang, Houwen Wu, and Shuxuan Ying.
\newblock {Validity of Thermodynamic Laws and Weak Cosmic Censorship for AdS
  Black Holes and Black Holes in a Cavity}.
\newblock 2 2020.
\newblock \href {http://arxiv.org/abs/2002.12233} {\path{arXiv:2002.12233}}.

\bibitem{McGough:2016lol}
Lauren McGough, Márk Mezei, and Herman Verlinde.
\newblock {Moving the CFT into the bulk with $ T\overline{T} $}.
\newblock {\em JHEP}, 04:010, 2018.
\newblock \href {http://arxiv.org/abs/1611.03470} {\path{arXiv:1611.03470}},
  \href {https://doi.org/10.1007/JHEP04(2018)010}
  {\path{doi:10.1007/JHEP04(2018)010}}.

\bibitem{Simovic:2018tdy}
Fil Simovic and Robert.B. Mann.
\newblock {Critical Phenomena of Charged de Sitter Black Holes in Cavities}.
\newblock {\em Class. Quant. Grav.}, 36(1):014002, 2019.
\newblock \href {http://arxiv.org/abs/1807.11875} {\path{arXiv:1807.11875}},
  \href {https://doi.org/10.1088/1361-6382/aaf445}
  {\path{doi:10.1088/1361-6382/aaf445}}.

\bibitem{Simovic:2019zgb}
Fil Simovic and Robert~B. Mann.
\newblock {Critical Phenomena of Born-Infeld-de Sitter Black Holes in
  Cavities}.
\newblock {\em JHEP}, 05:136, 2019.
\newblock \href {http://arxiv.org/abs/1904.04871} {\path{arXiv:1904.04871}},
  \href {https://doi.org/10.1007/JHEP05(2019)136}
  {\path{doi:10.1007/JHEP05(2019)136}}.

\bibitem{Haroon:2020vpr}
Sumarna Haroon, Robie~A. Hennigar, Robert~B. Mann, and Fil Simovic.
\newblock {Thermodynamics of Gauss-Bonnet-de Sitter Black Holes}.
\newblock {\em Phys. Rev. D}, 101:084051, 2020.
\newblock \href {http://arxiv.org/abs/2002.01567} {\path{arXiv:2002.01567}},
  \href {https://doi.org/10.1103/PhysRevD.101.084051}
  {\path{doi:10.1103/PhysRevD.101.084051}}.

\end{thebibliography}
\end{document}